\documentclass[12pt, appendixfloats, numberedappendix]{emulateapj}

\usepackage{hyperref}
\usepackage{color}
\usepackage{natbib}
\usepackage{rotating}
\usepackage{amsmath}

\newcommand{\beq}{\begin{equation}}
\newcommand{\eeq}{\end{equation}}

\def\mgii{\ion{Mg}{2}}

\def\mgiidoublet{\ion{Mg}{2}~$\lambda\lambda2796,2803$}

\def\feii{\ion{Fe}{2}}

\def\caii{\ion{Ca}{2}}

\def\ovi{\ion{O}{6}}

\begin{document}

\title{The properties of the cool circumgalactic gas\\ probed with the SDSS, WISE and GALEX surveys}
\shorttitle{galaxy-absorber correlations}

\shortauthors{T.W. Lan et al.}
\author{
Ting-Wen Lan\altaffilmark{1}, 
Brice M{\'e}nard\altaffilmark{1,2,3}, \&
Guangtun Zhu\altaffilmark{1}
} 
\altaffiltext{1}{Department of Physics \& Astronomy, Johns Hopkins University, 3400 N. Charles Street, Baltimore, MD 21218, USA, tlan@pha.jhu.edu}
\altaffiltext{2}{Kavli IPMU (WPI), the University of Tokyo, Kashiwa 277-8583, Japan}
\altaffiltext{3}{Alfred P. Sloan Fellow} 
 
\begin{abstract}
We explore the distribution of cool ($\sim10^4\,$K) gas around galaxies and its dependence on galaxy properties. By cross-correlating about 50,000 \mgii\ absorbers with millions of sources from the SDSS (optical), WISE (IR), and GALEX (UV) surveys we effectively extract about 2,000 galaxy-absorber pairs at $z\sim0.5$ and probe relations between absorption strength and galaxy type, impact parameter and azimuthal angle. We find that cool gas traced by \mgii\ absorbers exists around both star-forming and passive galaxies with a similar incidence rate on scales greater than {100}\,kpc but each galaxy type exhibits a different behavior on smaller scales: \mgii\ equivalent width does not correlate with the presence of passive galaxies whereas stronger \mgii\ absorbers tend to be found in the vicinity of star-forming galaxies. This effect is preferentially seen along the minor axis of these galaxies, suggesting that some of the gas is associated with outflowing material. In contrast, the distribution of cool gas around passive galaxies is consistent with being isotropic on the same scales. We quantify the average excess \mgii\ equivalent width $\langle \delta W_{0}^{\rm Mg\,II}\rangle$ as a function of galaxy properties and find $\langle\delta W_{0}^{\rm Mg\,II}\rangle \propto SFR^{1.2}$, $sSFR^{0.5}$ and $M_\ast^{0.4}$ for star-forming galaxies. This work demonstrates that the dichotomy between star-forming and passive galaxies is reflected in the circumgalactic medium traced by low-ionized gas. We also measure the covering fraction of \mgii\ absorption and find it to be about 2-10 times higher for star-forming galaxies than passive ones within 50\,kpc. We estimate the amount of neutral gas in the halo of $\langle \log M_{\ast}/{\rm M_\odot} \rangle \sim10.8$ galaxies to be a few $\times\, 10^9\,{\rm M_\odot}$ for both types of galaxies. Finally, we find that correlations between absorbers and sources detected in the UV and IR lead to physical trends consistent with those measured in the optical.
\end{abstract}


%% =======================================
\section{Introduction}
%% =======================================

The circum-galactic medium (CGM), tracing baryons surrounding galaxies within their own dark matter halos, plays an important role in galaxy formation and evolution. This interface between galaxies and the inter-galactic medium (IGM) is known to harbor gas flows as accretion and/or outflows but its properties and dependence on galaxy properties are poorly known. For several decades, metal absorption lines imprinted in the spectra of background sources have been the main tool to probe the gas distribution with the CGM. Studies have made use of galaxy-absorber pairs, for which the galaxy has been spectroscopically confirmed to be close to the redshift of the absorber. Various absorption lines can be used for such analyses. Among them the \mgiidoublet\ absorption lines, which trace cool gas (T $\sim10^{4}$ K), have been extensively used due to their strength and visibility from the ground: $0.3<z<2.5$ in the optical range.
From the first detection of a galaxy-\mgii\ absorber pair by \citet{Bergeron1986}, to the collection of a sample of about 50 systems by \cite{Steidel1994} to the latest extension to about 200 pairs by \citet{NielsenMgIIcatalogI}, numerous authors have attempted to find correlations between galaxy and absorber properties but the large scatter typically observed among such quantities has mostly led to non-detections \citep[e.g.][]{Kacprzak2011}. 
Another line of investigation has been based on statistical analyses of large surveys \citep[e.g.][]{Zibetti2007,MenardOII,BordoloiMgII}. While some interesting trends have emerged from these measurements, the connection between galaxies and the distribution of cool gas in the CGM is, after more than two decades of studies, still poorly constrained. To provide some guidance to the theoretical understanding of the CGM, additional observational constraints that can possibly reveal the physical connections between absorber and galaxy properties are needed. 

In this paper, we statistically extract about 2000 galaxy-absorber pairs using data from the Sloan Digital Sky Survey (SDSS; \citealt{YorkSDSS}). These pairs can then be used to measure correlations between absorber and galaxy properties such as star formation rate (SFR), stellar mass, and azimuthal angle as a function of impact parameter. We also perform our analysis to data from the GALEX and WISE surveys and provide additional support to our findings using UV and IR data. The outline of the paper is as follows: we present the datasets and statistical estimators in Section 2 and the results of galaxy-absorber correlations in Section 3. We discuss our findings in Section 4 and summarize our results in Section 5. In this analysis, we adopt flat $\Lambda$CDM cosmology with $h=0.7$ and $\Omega_{M}=0.3$. Throughout this work we use AB magnitude system and unless stated otherwise scales are in physical units.

%---------------------------------   
% figure 
%----------------------------------
\begin{figure*}
\begin{center}
\includegraphics[viewport=20 350 900 920,scale=0.45]{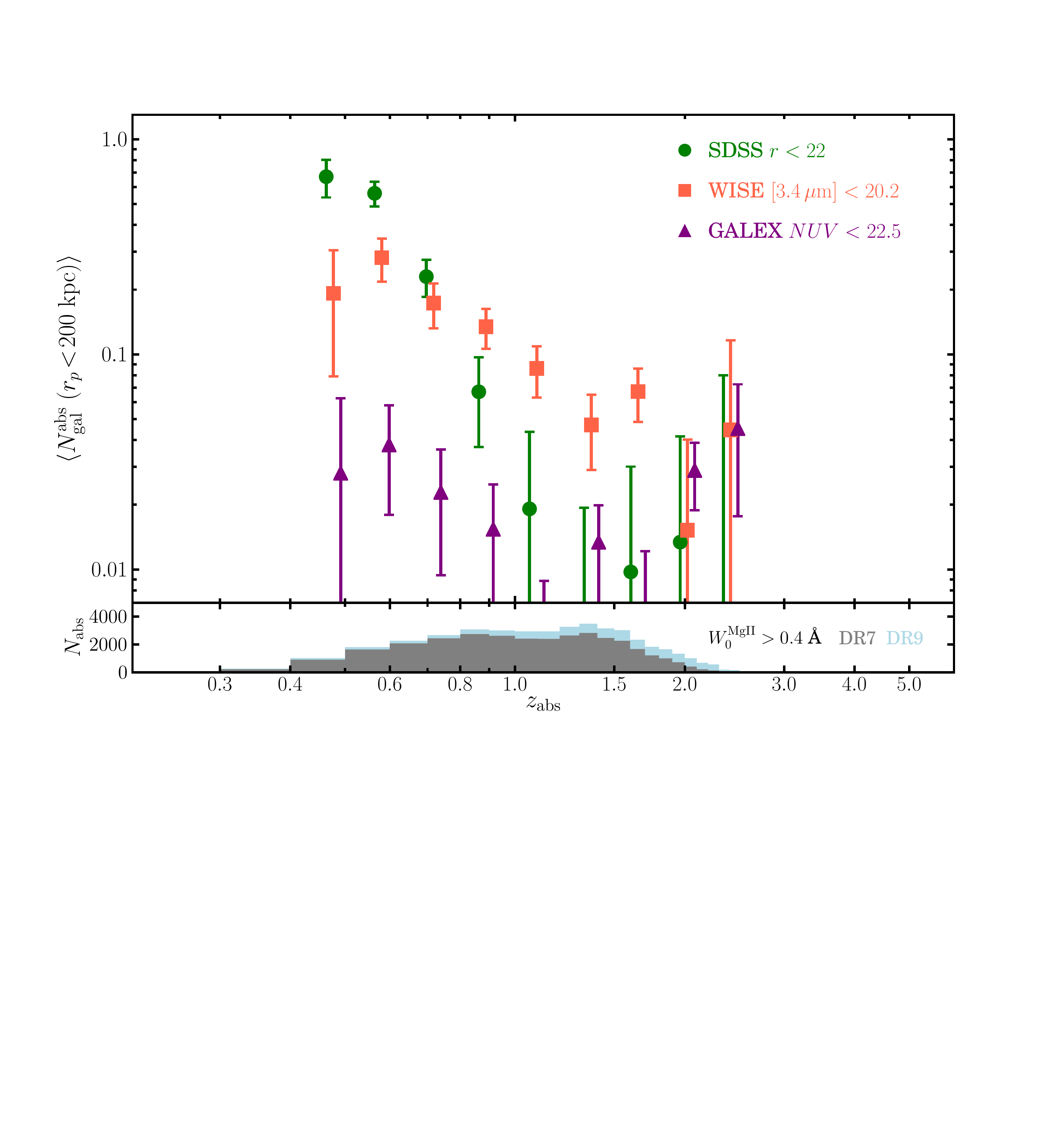}
\caption{\emph{Top}: SDSS, WISE, and GALEX mean number of galaxies per absorber as a function of $z_{\rm abs}$ from $0.4<z_{\rm abs}<2.8$.  We search galaxies with impact parameter from 10 to 200 kpc in SDSS and 50 kpc to 200 kpc in WISE and GALEX. Three surveys all detect galaxies associated with \mgii\ absorbers up to $z\sim1$. From redshift 0.4 to 0.6, SDSS detects 70\% of galaxies associated with \mgii\ absorbers, WISE detects 30\% and GALEX detects 4\%.
\emph{Bottom}: The redshift distributions of DR7 and DR9 \mgii\ absorbers.}
\label{fig:3surveys}
\end{center}
\end{figure*}
%----------------------------------

%%============================================================
\section{Data analysis}
\label{sec:datasets}
\subsection{Datasets}
%%============================================================

We use a sample of absorber systems from the JHU-SDSS \mgii/\feii\ absorption line catalog\footnote{\url{http://www.pha.jhu.edu/~gz323/jhusdss}} compiled by \citet{ZhuMgIIcatalog}. This sample was created using an automatic algorithm to detect absorption lines in quasar spectra. The authors detected about 36,000 and 11,000 \mgii\ absorbers in SDSS DR7 \citep{Schneiderdr7qso} and DR9 \citep{Parisdr9qso} quasar spectra, respectively. In this analysis we focus on systems with \mgii $\,\lambda 2796$ rest equivalent width $W_{0}^{\rm Mg\,II} >0.4$\,\AA\ \footnote{This absorption strength limit corresponds to conventional strong absorbers.} which include about 90\% of the systems in the original catalog. The redshift distributions of the selected systems are shown in the bottom panel of Figure~\ref{fig:3surveys}.

To select galaxies, we use catalogs from
\begin{itemize}
\item the SDSS DR7 photometric catalog \citep{AbazajianDR7}. We select galaxies with dereddened magnitudes in the range $18<r<22$. With an average angular resolution of about $1.4\arcsec$, SDSS allows us to probe quasar-galaxy pairs on impact parameters greater than about\,10 kpc at $z=0.5$. The magnitudes reported in this work are apparent magnitudes.

\item the \textit{Wide-field Infrared Survey Explorer} (WISE; \citealt{WrightWISE}). We make use of the all-sky data release\footnote{\url{http://irsa.ipac.caltech.edu/Missions/wise.html}} and select objects brighter than 20.2 in W1 [$3.4\,\mu$m] magnitude, which is close to the flux limit. The average angular resolution of WISE [$3.4\,\mu$m] band is about $6\arcsec$, corresponding to about 40\,kpc at $z=0.5$.
\item the \textit{Galaxy Evolution Explorer} (GALEX; \citealt{MartinGALEX}) all sky imaging survey (AIS). We use sources from the Data release DR4/DR5\footnote{\url{http://galex.stsci.edu/}} \citep{GALEX_ref} with NUV (1770-2730\,\AA) brighter than 22.5 mag. Similar to WISE, the average angular resolution of NUV band is about $6 \arcsec$. 
\end{itemize}

%=========================================
\subsection{Method}
\label{method}

We measure correlations between the presence of absorbers detected in quasar spectra and photometric galaxies (without using any galaxy redshift information) detected within some angular aperture. By applying a statistical method to subtract galaxies not associated with absorbers, we are able to study correlations between galaxies and absorbers. In the following, we introduce two estimators to extract such correlations: (1) we measure the mean number of galaxies associated with selected absorbers, as a function of galaxy properties and (2) we estimate the change in absorber equivalent width as a function of proximity to a given type of galaxies.

\subsubsection{The mean number of galaxies per absorber}

We statistically extract galaxies associated with absorbers by comparing the number of galaxies
observed along absorber lines of sight with the number of galaxies measured at random positions in the sky.
%around quasars with absorbers with the mean number of galaxies on the sky. 
A difference is expected to arise from the presence of galaxies associated with absorbers. 
%To do so we first define the number of galaxies observed around the line of sight of a quasar $Q$ as $N_{\rm gal}^{Q}$. 
%To do so, 
For each intervening absorber,
%detected in the spectrum of a quasar, 
the excess in the number of galaxies found within a given angular aperture is given by
%we measure  excess in the number of galaxies around For each
%this line of sight with respect to the mean value on the sky estimated from reference quasars.} This can be written as
\begin{equation}
\delta N^{{\rm abs},j}_{\rm gal} = N_{\rm gal}^{{\rm Q(abs},j)} - \langle N_{\rm gal}^{\rm Q(ref)} \rangle\;,
\label{eq:N_abs}
\end{equation}
where $N_{\rm gal}^{{\rm Q(abs},j)}$ is the number of galaxies (including both associated galaxies and background/foreground galaxies) around a quasar with detected absorber j and
$\langle N_{\rm gal}^{\rm Q(ref)} \rangle$ is the mean number of galaxies around reference quasars (the mean number of galaxies on the sky). 
We use reference quasars as opposed to random positions to also subtract a potential contribution arising from galaxies clustered with the quasars.
To precisely construct a set of reference quasars, for each quasar with an absorber we randomly select four quasars with similar quasar redshift and i-band magnitude with $|\Delta z|<0.05$ and $|\Delta i|<0.05$ mag as reference quasars. Such selection is done without any prior knowledge of the existence of absorbers along line-of-sights. When fewer quasars satisfy these criteria, we increase the search radius in both redshift and magnitude to find matching objects. Once the reference quasars are selected, all galaxies around the absorber and reference quasars are assigned to have the same redshift of the absorber.
The quantity $\delta N^{\rm abs,{\it j}}_{\rm gal}$ can be positive or negative along a given line of sight due to Poisson fluctuations dominated by the distributions of foreground and/or background galaxies.

By averaging the excess number of galaxies (Eq. \ref{eq:N_abs}) over an ensemble of absorbers, we obtain the mean number of galaxies associated with a \mgii\ absorber $\langle N_{\rm gal}^{\rm abs} \rangle$:
%(expected to be positive) 
%as
%, which is expected to be a positive quantity, is obtained by averaging the above quantity for an ensemble of absorbers, taking into account potential incompleteness effects due to the blending between quasars and galaxies at small angular separation ($<4\,\arcsec$) when counting galaxies:
\begin{equation}
\langle N_{\rm gal}^{\rm abs} \rangle =\langle \delta N^{{\rm abs},j}_{\rm gal}\;\times w(\theta_j,C_j) \rangle\;,
\label{main}
\end{equation}
where $w(\theta,C)$ is a correction function accounting for blending effects between quasars and galaxies at small angular separation when counting galaxies, $\theta$ is the angular separation  between quasars and galaxies and C is the galaxy $g-i$ color.
The detail of this completeness correction function $w(\theta,C)$ is presented in Appendix A. 
This correction is only important when $\theta<4\arcsec$.
We estimate the error on $\langle N_{\rm gal}^{\rm abs} \rangle$ with Poisson statistics (we note that the counts are usually dominated by the contribution from uncorrelated background objects) and we primarily focus on scales smaller than 200\,kpc ($\sim30\,\arcsec$ at $z=0.5$) to probe galaxies directly associated with absorbers. This method enables us to statistically probe the properties of galaxies associated with absorbers without matching the exact galaxy-absorber pairs or assigning the properties of a certain galaxy to an absorber. It includes galaxies in all types of environments (isolated and in groups). 
If we assume that each absorber is connected to one galaxy, the mean number of galaxies associated with \mgii\ absorbers $\langle N_{\rm gal}^{\rm abs} \rangle$ represents the fraction of galaxies associated with absorbers detected within the flux limit of the survey. The quantity $\langle N_{\rm gal}^{\rm abs} \rangle$ can be measured as a function of various parameters such as impact parameter, absorption strength, and galaxy properties (Section~\ref{galaxy_type_mgii}) to provide us with insights into the galaxy-absorber connection.
%----------------------------------
% figure
%----------------------------------
\begin{figure*}
\begin{center}
\includegraphics[scale=0.45]{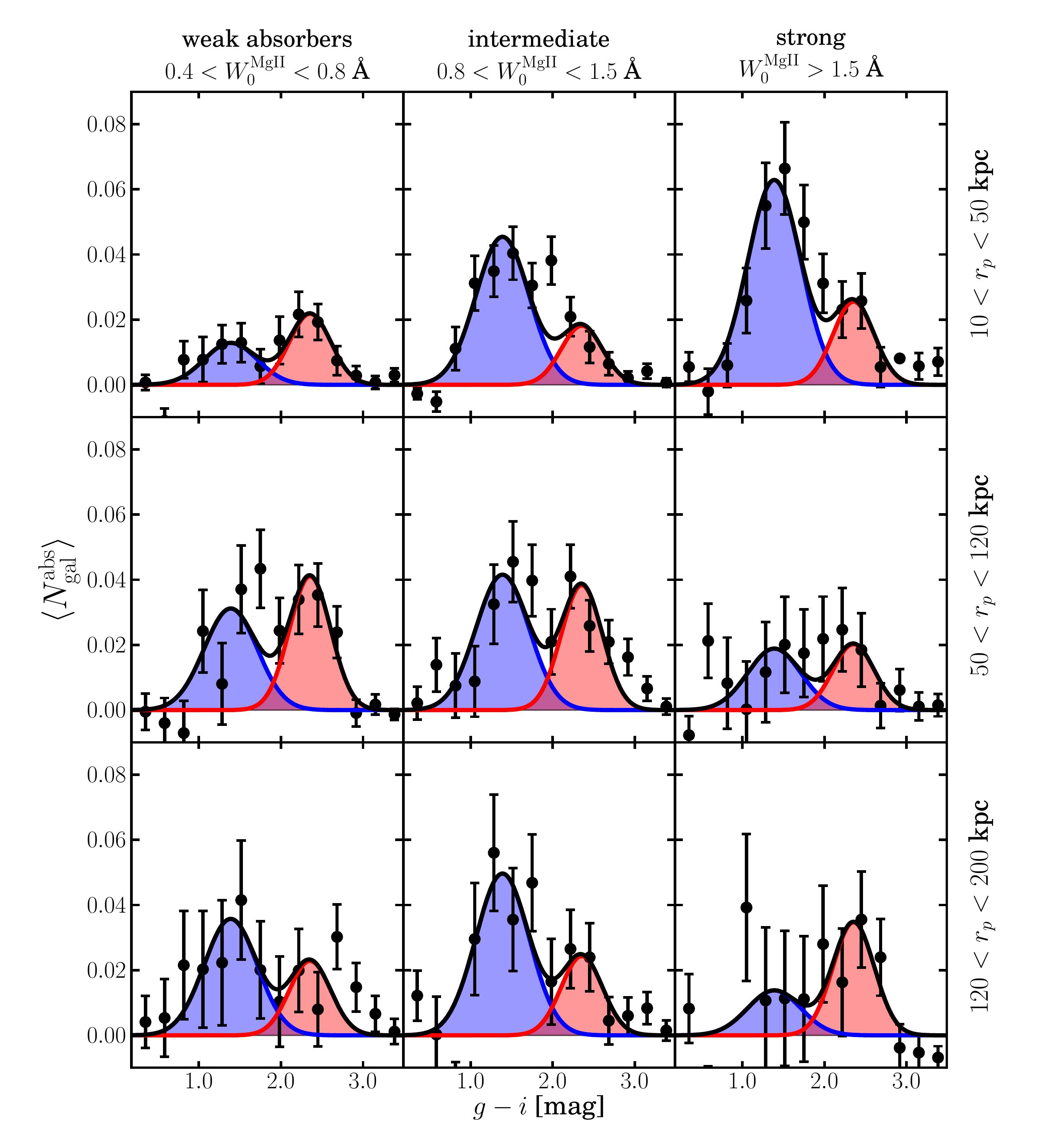}
\caption{The mean number of SDSS galaxies per absorber as a function of g-i color, $W_{0}^{\rm Mg\,II}$, and impact parameter with $0.4<z<0.6$. The blue and red shaded regions are the best-fit Gaussians, for which the centers and the widths are fixed values derived from PRIMUS galaxies. The most significant trend occurs at small impact parameters where the fraction of blue galaxies increases significantly as $W_{0}^{\rm Mg\,II}$ increases.}
\label{fig:SDSS_color_distribution}
\end{center}
\end{figure*}
%----------------------------------

%----------------------------------
% figure
%----------------------------------
\begin{figure*}
\begin{center}
\includegraphics[scale=0.5]{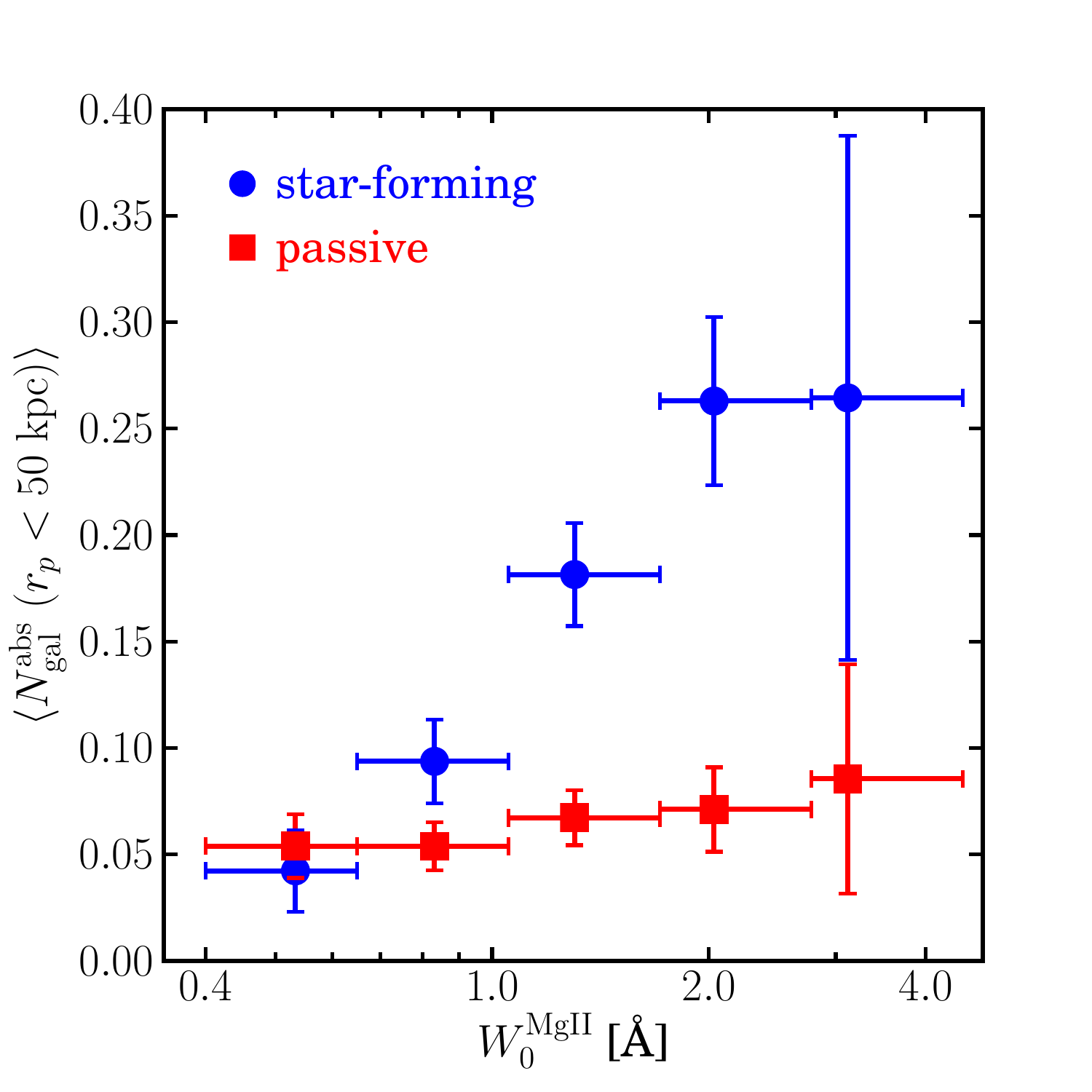}
\includegraphics[scale=0.5]{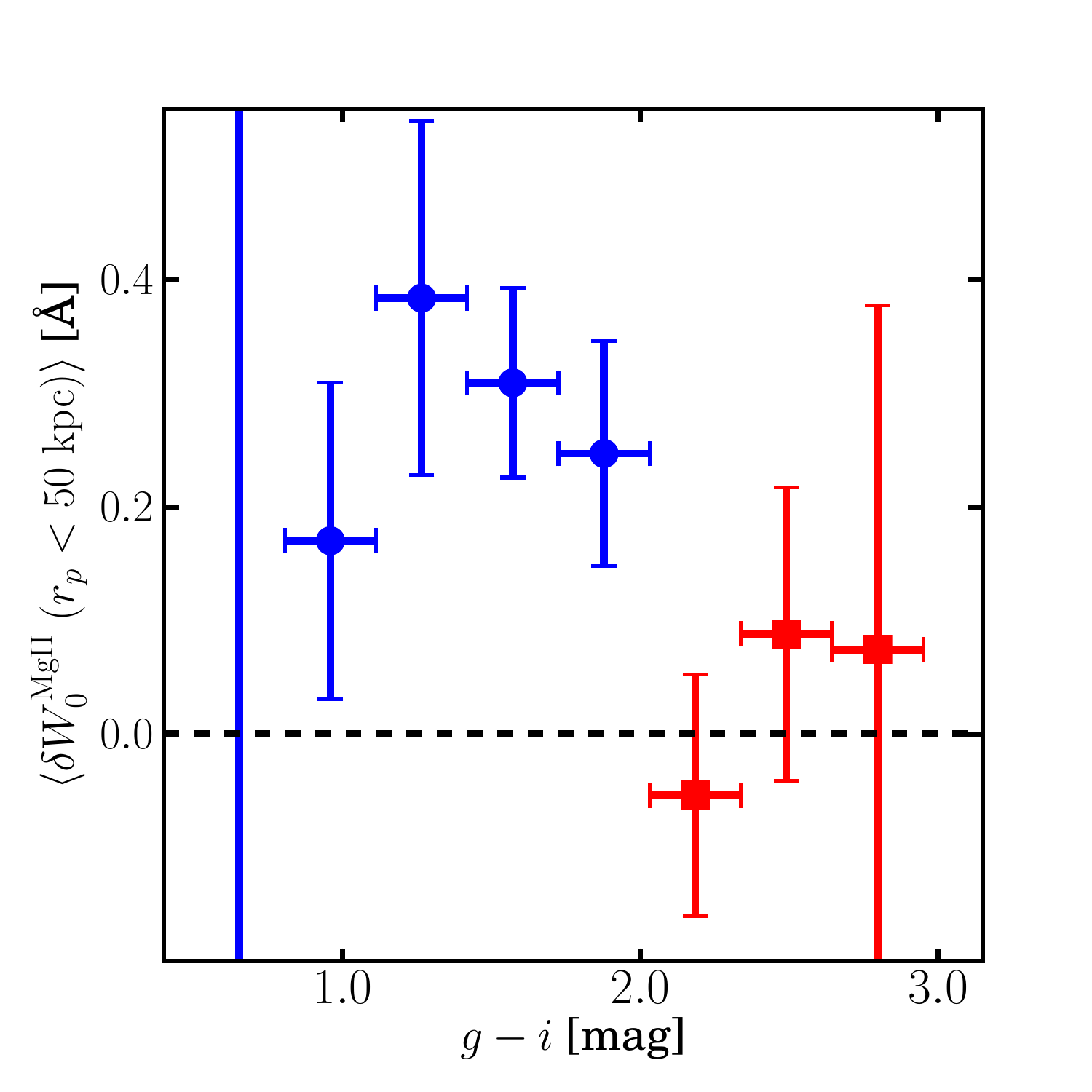}
\caption{\emph{Left:} The mean number of star-forming and passive galaxies per absorber at small impact parameter ($10<r_{p}<50$ kpc). 
The horizontal bars show the range of $W_{0}^{\rm MgII}$ values in each bin and the data point abscissae indicate the median $W_{0}^{\rm MgII}$ value within the bin.
We apply color cuts to separate galaxies into two components: star-forming ($0.5<g-i<2.0$), and passive ($2.0<g-i<3.0$). The number of star-forming galaxies increases significantly as a function of $W_{0}^{\rm MgII}$. On the other hand, the number of passive galaxies remains consistent. \emph{Right:} The excess absorption as a function of galaxy g-i color. The mean absorption strength around star-forming galaxies is about 0.3 \AA \ higher than the mean absorption strength of the sample, while such an excess is not seen for passive galaxies. The left blue data point has value about -0.3 in y-axis which is beyond the range of our plot.}
\label{fig:small_scale_correlation}
\end{center}
\end{figure*}
%----------------------------------

\subsubsection{The excess \mgii\ absorption around galaxies}

We estimate the excess \mgii\ absorption strength with respect to the mean of the absorber sample around a given type of galaxies to investigate how the presence of a given type of galaxies influences the amount of detectable \mgii\ absorption. To do so we correlate the rest equivalent width of each \mgii\ absorber with the number of associated galaxies around it. The relative excess absorption strength around a given type of galaxies is defined as 
\begin{equation}
\langle \delta W_{0}^{\rm Mg\,II} \rangle_{\rm gal} = \langle W_{0}^{\rm Mg\,II} \rangle_{\rm gal} - \langle W_{0}^{\rm Mg\,II} \rangle_{\rm sample}\,,
\label{excess_absorption}
\end{equation}
where $\langle W_{0}^{\rm Mg\,II} \rangle_{\rm sample}$ is the mean rest equivalent width of the absorber sample, which is equal to 1.14 \AA, and $\langle W_{0}^{\rm Mg\,II} \rangle_{\rm gal}$ is the mean rest equivalent width around a given type of galaxies which is estimated as
\begin{equation}
\langle W_{0}^{{\rm Mg\,II}} \rangle_{{\rm gal}} =
\frac{
{\displaystyle
\sum_{j=1}^{\rm N_{abs}} W_{0, {\it j}}^{\rm Mg\,II}\,
 \delta N_{\rm gal}^{{\rm abs},j}\;
 w(\theta_j,C_j)
}
}
{
{\displaystyle
\sum_{j=1}^{{\rm N_{abs}}} \delta N_{\rm gal}^{{\rm abs},j}\,
 w(\theta_j,C_j)
}
}
\end{equation}
with $N_{\rm abs}$ being the total number of absorbers, $W^{\rm Mg\,II}_{0,{\it j}}$ being the rest equivalent width of an absorber j, $\delta N_{\rm gal}^{{\rm abs},j}$ and $w(\theta_{j},C_{j})$ are given in Eq.~\ref{eq:N_abs} and \ref{main}. 
The excess rest equivalent width around galaxies $\langle \delta W_{0}^{\rm Mg\,II} \rangle_{\rm gal}$ can be investigated as a function of galaxy properties and impact parameter. When doing so we note that the mean absorption strength of the absorber sample  $\langle W_{0}^{\rm Mg\,II} \rangle_{\rm sample}$ remains the same. This will be presented in Section \ref{sec:excess_absorption}.

%% ============================================================
\section{Results}
We first measure the mean number of galaxies per absorber $\langle N_{\rm gal}^{\rm abs}\rangle$ defined in Eq.~\ref{main} for the three photometric surveys presented in Section~\ref{sec:datasets}, as a function of absorber redshift. This allows us to probe the redshift range over which each survey can detect galaxies associated with the selected \mgii\ absorbers. In each case we use a search radius ranging from the size of the angular resolution up to a projected radius of 200 kpc. The results are presented in Figure~\ref{fig:3surveys}. SDSS, WISE and GALEX all detect galaxies associated with \mgii\ absorbers with a wide range of redshift. At $z\sim0.5$ about $70\%$ of galaxies associated with our \mgii\ absorbers are detected in SDSS in the selected magnitude range. This fraction decreases to 30\% for WISE and about 4\% for GALEX. At $z>0.8$ we find more associations between absorbers and photometric objects in WISE than SDSS, showing that the connection between galaxies and absorbers can be studied with WISE at relative high redshifts. We first study galaxy-absorber pairs found in the SDSS in Section \ref{galaxy_type_mgii}. In Section \ref{sec:GALEX_WISE} we will perform our analysis to the GALEX and WISE surveys.

%% ==============================
\subsection{SDSS galaxy-absorber correlations}

\subsubsection{Galaxy type and spatial distribution}
\label{galaxy_type_mgii}

Focusing first on galaxies optically-selected from SDSS, we limit our analysis to the redshift range $0.4<z<0.6$ where the detection rate of galaxy-absorber pairs is the highest. Within 200 kpc, the mean number of SDSS galaxies around quasars with absorbers $\langle N_{\rm gal}^{\rm Q(abs)}\rangle$ is about 3.1 and the mean number of galaxies around reference quasars $\langle N_{\rm gal}^{\rm Q(ref)}\rangle$  is about 2.4. This excess corresponds to an effective number of about 2000 galaxy-absorber pairs from SDSS photometric data. This is about ten times larger than any existing galaxy-absorber pair catalog (e.g. \citealt{NielsenMgIIcatalogI}). 

Before investigating the types of galaxies connected to \mgii\ absorbers, it is useful to characterize the properties of galaxies in the redshift range of interest. At $z\sim0.5$ the PRIsm MUlti-object Survey (PRIMUS; \citealt{CoilPRIMUS,CoolPRIMUS}) provides us with a large and complete reference galaxy sample for which both broadband colors and low-resolution spectroscopic redshifts are available. We use the first data release\footnote{\url{http://primus.ucsd.edu/}.} of the survey and measure the distribution of $g-i$ color for galaxies with redshift $0.4<z<0.6$ and brightness $18<r<22$. This distribution is well represented by the sum of two Gaussians, one centered at $g-i=1.39$ with width of $0.32$ mag and the other one centered at $2.35$ with a width of $0.25$ mag. The relative amplitude between the blue and red is 1.4 to 1.0.

Considering the distribution of PRIMUS galaxies representative of the galaxy population at $z\sim0.5$, we now investigate the properties of galaxies found to lie next to \mgii\ absorbers. We measure $\langle N_{\rm gal}^{\rm abs}\rangle$ as a function of \mgii\ absorption strength, galaxy color and impact parameter. The results are shown in Figure~\ref{fig:SDSS_color_distribution}. The solid lines show fits of the data points using the same pair of Gaussians used to characterize the color distribution of galaxies in PRIMUS (we fix the centers and the widths, as shown with the blue and red shaded regions). Our analysis shows that \mgii\ absorbers exist around both blue and red galaxies. Figure \ref{fig:SDSS_color_distribution} shows a number of trends:
\begin{itemize}
\item Galaxies associated with weak absorbers (left column, $0.4<W^{\rm Mg\,II}_{0}<0.8$\,\AA) are mostly found within intermediate ($50<r_{p}<120\,$kpc) and large ($120<r_p<200\,$kpc) impact parameters, while galaxies associated with strong absorbers (right column,  $W^{\rm Mg\,II}_{0}>1.5$\,\AA) tend to be within small impact parameters ($10<r_p<50\,$kpc). 
\item For weak absorbers (left column), galaxies show a bimodal distribution of colors which does not strongly depend on impact parameter. In contrast, the majority of galaxies associated with strong absorbers appear to be blue galaxies with small impact parameters. Galaxies associated with intermediate absorbers (middle column,  $0.8<W^{\rm Mg\,II}_{0}<1.5$\,\AA) have mixed behaviors. 
\item At small impact parameters, our results show that the number of blue galaxies increases significantly as $W^{\rm Mg\,II}_{0}$ increases, showing a correlation between strong absorbers and star-forming galaxies. 
\end{itemize}

To further study this correlation, we separate the galaxies according to color: 
\begin{eqnarray}
\text{star-forming: }& 0.5<g-i<2.0\nonumber \\
\text{passive: }& 2.0<g-i<3.0\;.
\label{eq:colorcut}
\end{eqnarray}
Using this separation, we show in the left panel of Figure~\ref{fig:small_scale_correlation} the mean number of star-forming and passive galaxies as a function of \mgii\ equivalent width for $10<r_p<50$\,kpc. 
This dependence on absorber equivalent width reveals two distinct trends: over the range $0.4<W_{0}^{\rm Mg\,II}<4.5\,$\AA, we observe the incidence rate of star-forming galaxies to increase by about a factor of six, indicating that stronger absorbers are more connected to star-forming galaxies within 50\,kpc. On the other hand, the mean number of passive galaxies found within 50\,kpc of absorbers appears to be independent on \mgii\ absorption strength. While our results show that \mgii\ absorbers live around both types of galaxies, their equivalent width appears to be primarily a function of the proximity to star-forming galaxies.

%----------------------------------
\begin{figure*}[t]
\begin{center}
\includegraphics[scale=0.52]{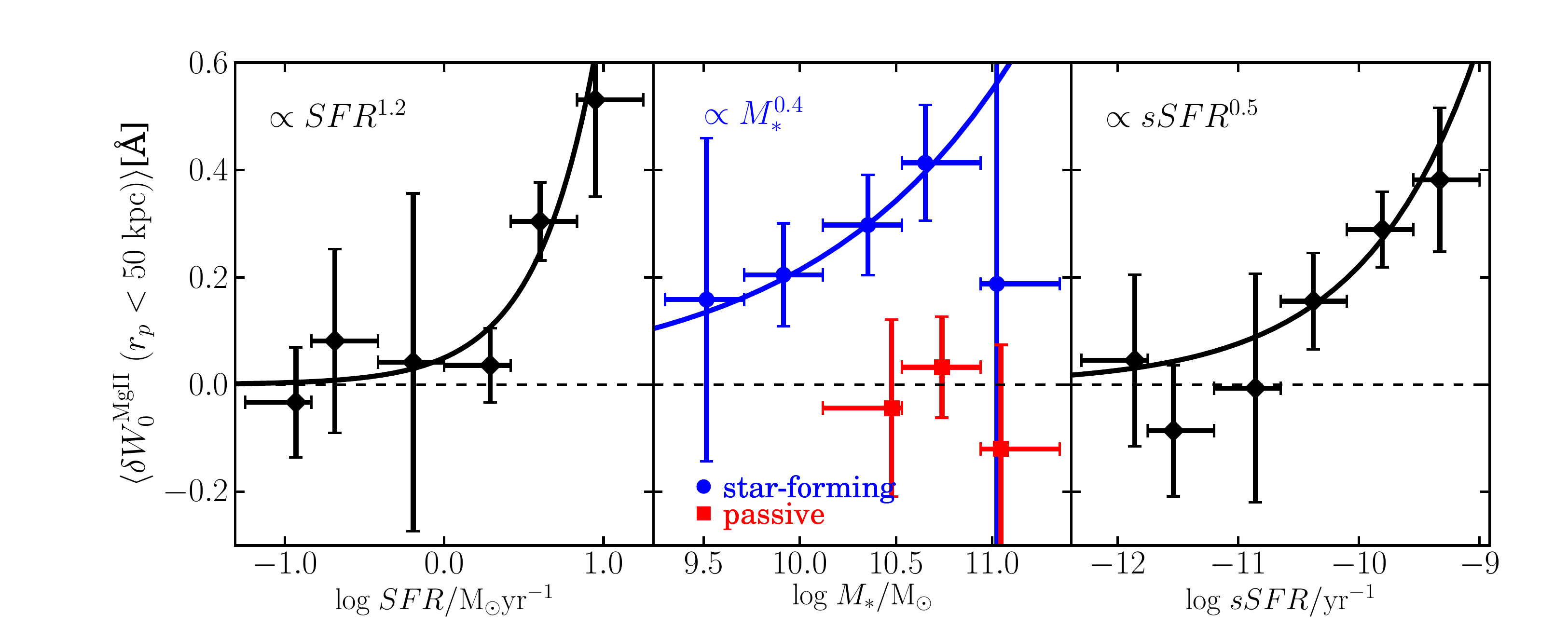}
\caption{The excess absorption as a function of physical properties of galaxies. The solid lines are the best-fit power laws shown in the top-left corner of each panel.  \emph{Left}: The excess absorption as a function of SFR. \emph{Middle}: The excess absorption as a function of stellar mass with color cuts separating star-forming and passive galaxies. \emph{Right}: The excess absorption as a function of sSFR.  The horizontal bars show the selected bin range and the data point abscissae indicate the median value within each bin.}
\label{fig:small_scale_cross_correlation}
\end{center}
\end{figure*}
%----------------------------------

%% =========================
\subsubsection{The excess absorption as a function of galaxy type}
\label{sec:excess_absorption}

We now investigate the galaxy-absorber connection considering the excess absorption correlated with the presence of a given type of galaxies, as defined in Equation~\ref{excess_absorption}. The results are shown in the right panel of Figure~\ref{fig:small_scale_correlation} where we have used bootstrap re-sampling to estimate the errors. The measurement clearly shows that the presence of galaxies with blue $g-i$ colors within 50\,kpc of \mgii\ absorbers increases the mean absorption equivalent width by about 0.3\,\AA. This amplitude reflects the \emph{mean} correlation, i.e. averaged over all impact parameters within 50 kpc and all galaxy orientations. 
If a fraction $f$ of absorbers around galaxies is responsible for this correlation, the excess equivalent width associated with these systems will intrinsically be higher by a factor $1/f$. This could be due, for example, to a preferred distribution of such systems within a given solid angle around galaxies  (as will be shown in Section~\ref{sec:angle}).
As shown in Figure~\ref{fig:small_scale_correlation}, this excess absorption drops for galaxies with $g-i>2.0$ and becomes consistent with zero. The transition takes place exactly at the color value separating star-forming and passive galaxies in the redshift range of interest. The dichotomy between star-forming and passive galaxies is therefore reflected in their cool, $\sim10^4\,$K gaseous halos on 50\,kpc scales.

%----------------------------------
\begin{figure}[h]
\begin{center}
\includegraphics[scale=0.45]{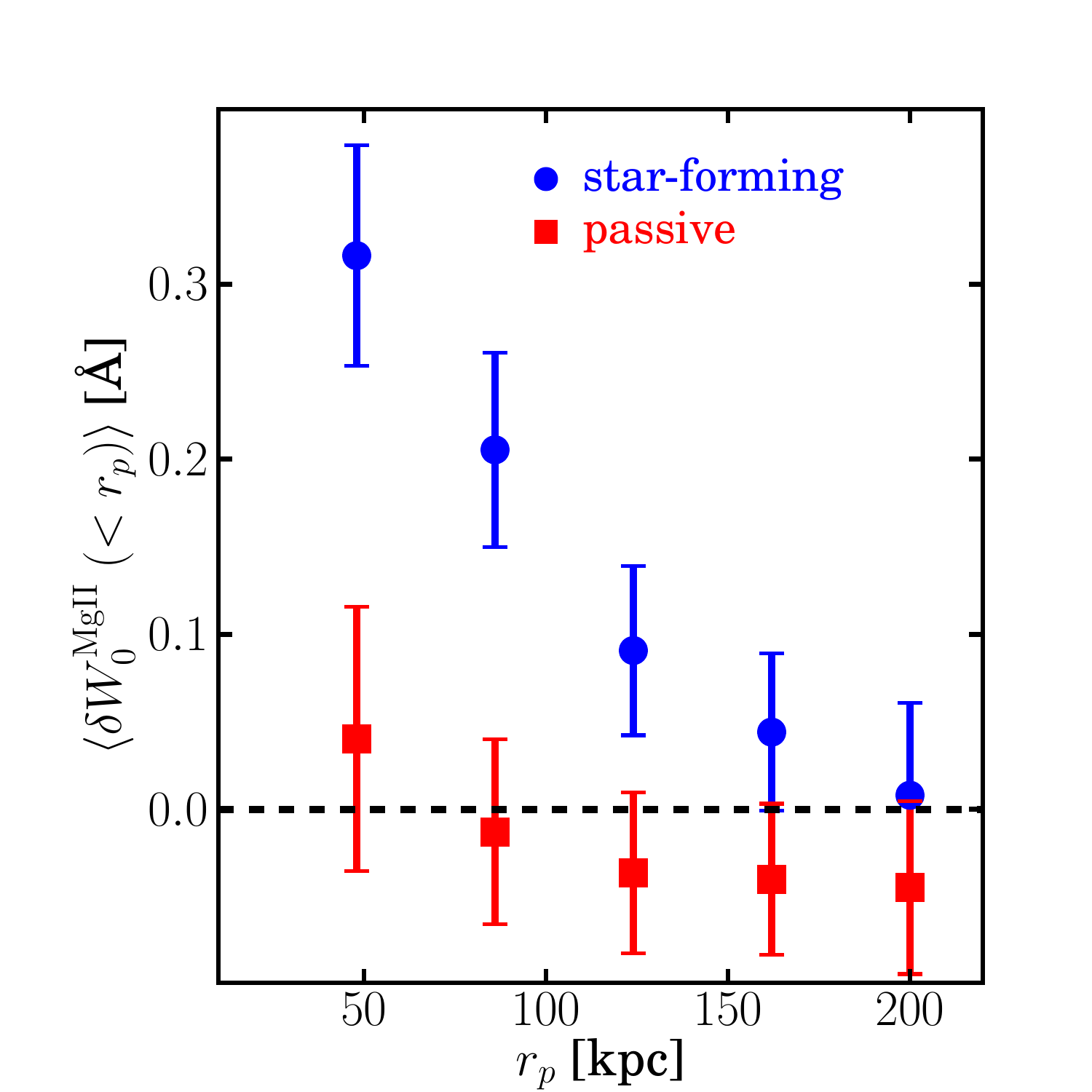}
\caption{Excess absorption measured within a given impact parameter. Note that the measurements are strongly correlated. For passive galaxies, the excess absorption is consistent with zero. On the other hand, the excess absorption of star-forming galaxies decreases when averaging over a larger area. The value of data points in the x-axis represents the maximum impact parameter of each bin.}
%, indicating that the correlation between star-forming galaxies and \mgii\ absorbers occurs at small scales. }
\label{fig:excess_scale}
\end{center}
\end{figure}

%% ----------------------------------
\begin{figure*}[]
\begin{center}
\includegraphics[scale=0.5]{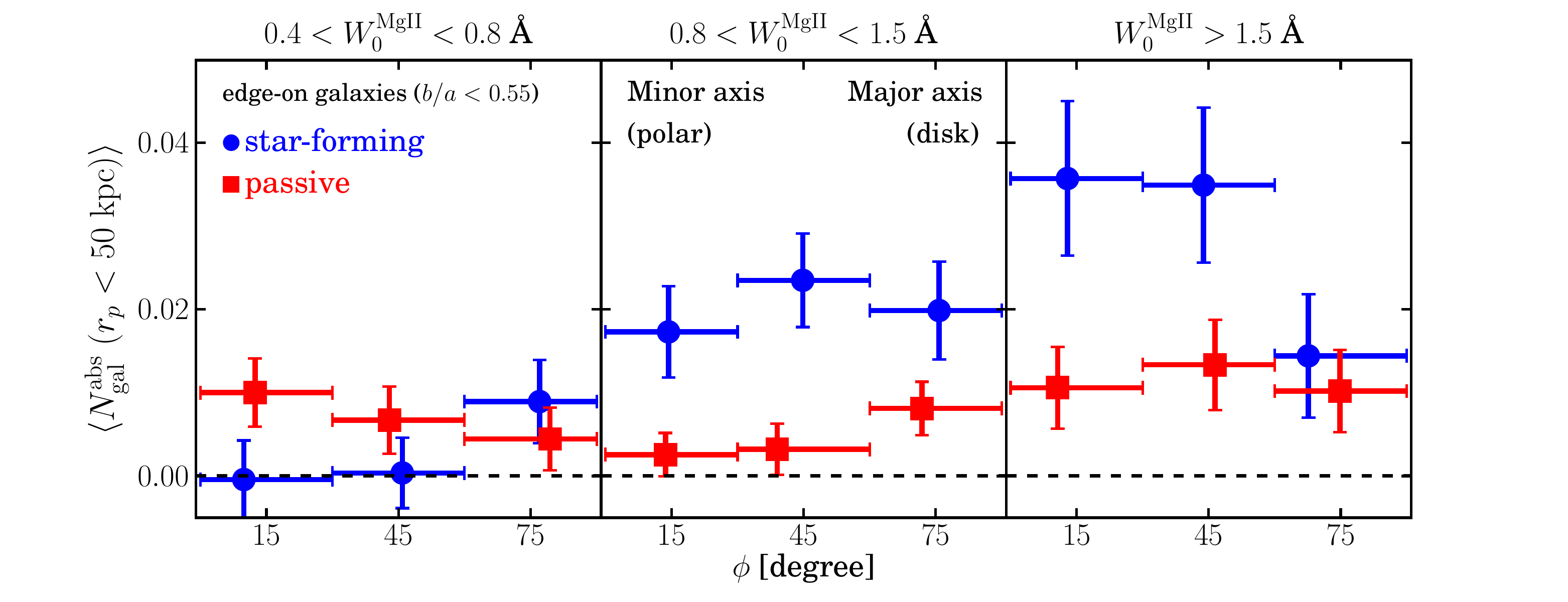}
\caption{The azimuthal angle dependence of \mgii\ absorbers and edge-on galaxies at $10<r_{p}<50$\,kpc. We select edge-on galaxies with axis ratio $b/a<0.55$ and separate star-forming ($0.5<g-i<2.0$) and passive ($2.0<g-i<3.0$) galaxies. A trend shows that strong absorbers preferentially exist near minor axis of star-forming galaxies. The blending correction is not applied in this analysis. The horizontal bars show the range of azimuthal angle in each bin and the data points indicate the median value within the bin.}
\label{fig:azimuthal_angle_dependence}
\end{center}
\end{figure*}

%%----------------------------------
\begin{figure*}
\begin{center}
\includegraphics[scale=0.5]{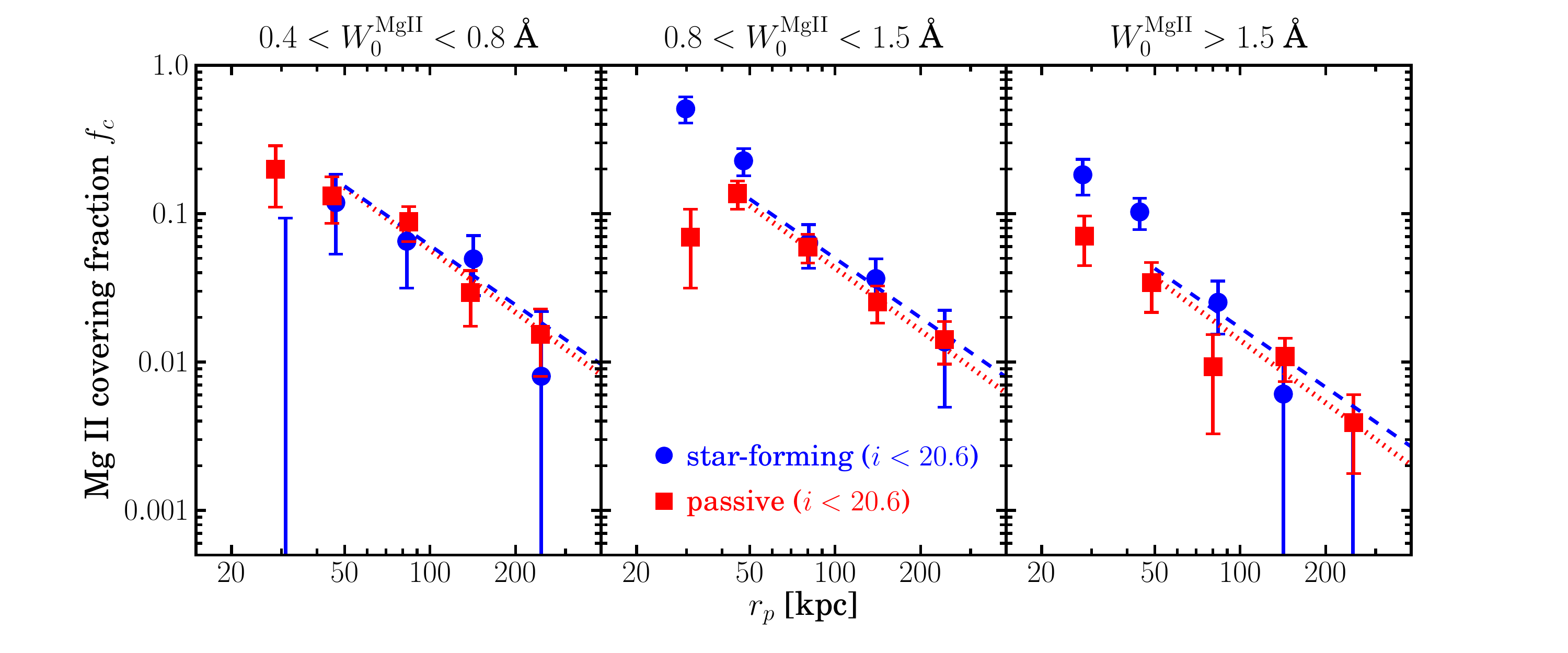}
\caption{The \mgii\ covering fraction of bright galaxies ($i<20.6$) as a function of $W_{0}^{\rm Mg\,II}$ from $20<r_{\rm p}<300\,\rm kpc$. For intermediate and strong absorbers, bright star-forming galaxies ($0.5<g-i<2.0$) have more than 2 times higher covering fraction than bright passive galaxies ($2.0<g-i<3.0$) within 50\,kpc, while \mgii\ covering fraction beyond 50\,kpc around both star-forming and passive galaxies decreases with a dependence of $f_{c}\propto r_{p}^{\alpha}$ with best-fit power laws shown in blue dashed lines and red dotted lines.}
\label{fig:covering_fraction_binning}
\end{center}
\end{figure*}
%----------------------------------

We now investigate how the excess absorption depends on physical properties of galaxies. To do so we estimate the average SFR and stellar mass of our galaxies using broadband photometry \citep[e.g.][]{Mostek2012}, and PRIMUS galaxies at $0.4<z<0.6$ with estimated physical properties from \citet{MoustakasPRIMUS} as a reference. Given a galaxy with $g-i$ color and $i$ band magnitude in our sample, we search for PRIMUS galaxies with similar observed properties $|\Delta (g-i)|<0.05$ and $|\Delta i|<0.05$, and use the average PRIMUS star formation rate and stellar mass values for these objects as an estimate of the parameters of the matched SDSS galaxy. When fewer than 10 PRIMUS galaxies satisfy these criteria, we increase the search radius to $|\Delta (g-i)|<0.1$ and $|\Delta i|<0.1$. We exclude a small subset of SDSS galaxies which do not have enough matches in PRIMUS. We note that these galaxies tend to be either too bright or too blue to be galaxies at $z\sim0.5$. We have verified that our results are not sensitive to the exact values of the matching criteria described above.

In Figure~\ref{fig:small_scale_cross_correlation} we show how the mean \mgii\ rest equivalent width changes as a function of SFR, stellar mass, and specific star formation rate (sSFR). As previously discussed, star-formation activity is correlated with excess \mgii\ absorption within 50\,kpc. Such a correlation is detected with $5\sigma$ for $\rm{SFR}\gtrsim1\,M_\odot/yr$ which is roughly the value differentiating star-forming from passive galaxies. To quantify this relation we fit the observed correlation, measured over about two and half orders of magnitude in SFR: $-1.2<\log SFR/{\rm M_\odot{\rm yr}^{-1}}<1.2$, with a power law function and find
\begin{eqnarray}
\langle \delta W_{0}^{\rm Mg\,II} \rangle &\propto & SFR^\alpha\, ~\text{with } \alpha=1.2\pm0.4\;.
\end{eqnarray}
This relation is shown as the solid line in the figure. The middle panel shows how the excess absorption depends on stellar mass. Here we find two distinct trends: a positive correlation between absorption strength and stellar mass for star-forming galaxies and no correlation for passive galaxies. Similarly to what was done above, we quantify this relation by fitting the observed trend with a power law function and find
\begin{eqnarray}
\langle \delta W_{0}^{\rm Mg\,II} \rangle&\propto& M_\ast^\beta  ~\text{with } \beta=0.4\pm0.3\
\end{eqnarray}
for star-forming galaxies and 
\begin{eqnarray}
\langle \delta W_{0}^{\rm Mg\,II} \rangle &\sim& 0  \text{~for passive galaxies. }
\end{eqnarray}
Finally we measure the correlation between absorption equivalent width and specific star formation rate ${\rm sSFR}=SFR/M_\ast$. This is shown in the right panel of the figure. We detect a positive correlation for galaxies with $\log sSFR / {\rm yr^{-1}} > -11$. A power-law fit, measured over three orders of magnitude in sSFR, gives

\begin{eqnarray}
\langle \delta W_{0}^{\rm MgII} \rangle &\propto & sSFR^\gamma\, {\rm with~} \gamma=0.5\pm0.2\;.
\end{eqnarray}

The above correlations are all reported for impact parameters smaller than 50\,kpc. They weaken and become consistent with zero on larger scales. This is illustrated in Figure~\ref{fig:excess_scale} where we show the excess absorption \emph{within} a given radius. Note that the measurements are therefore correlated. While the excess absorption gradually decreases for star-forming galaxies as a function of impact parameter, it is always consistent with zero for passive galaxies. The above relations present a set of observational constraints for theoretical models of gas accretion and outflows around galaxies.

%% =========================================
\subsubsection{Azimuthal angle dependence}
\label{sec:angle}

We now explore the azimuthal angle dependence of the correlation between galaxies and absorbers. To do so we make use of the minor-to-major axis ratio ($b/a$) and the position angle of galaxies retrieved from SDSS skyserver\footnote{\url{http://cas.sdss.org/dr7/en/}}. We select edge-on galaxies (with axis ratio $b/a<0.55$) as a function of their azimuthal angle with respect to the quasar line-of-sight and measure the mean number of galaxies found above the background, within an aperture of 50\,kpc.
The results are shown in Figure~\ref{fig:azimuthal_angle_dependence}\footnote{It should be noted that such a measurement does not require a precise position angle estimate for individual objects. It only depends on the correlation between azimuthal angle and the presence of an absorber. This is similar to the correlations measured in galaxy-galaxy lensing analyses.}. 

Passive galaxies do not show any preferred orientation for the whole range of $W_{0}^{\rm MgII}$. On the other hand, we find that the incidence rate of star-forming galaxies around absorbers depends on absorption strength and orientation. At $W_{0}^{\rm MgII}>1.5$\,\AA, we find about two times more systems perpendicular to the plane than along the direction of the disk. This trend is only observed at small impact parameters, i.e., with $r_p\lesssim 50$\,kpc. This geometry suggests that strong absorbers might be related to outflow gas produced from star-forming galaxies. Our results are consistent with those of \citet{BordoloiMgII}, \citet{BoucheMgII} and \citet{KacprzakAZ}, and also the stacking measurement of the \caii\ absorption at $z\sim0.1$ by \citet{ZhuCaII}.

% table
\begin{table*}[] 
\caption{The best fit parameters of \mgii\ covering fraction at $r_{p}>50$ kpc} %title of the table 
\centering % centering table 
\begin{tabular}{c cccc} % creating eight columns 
\hline\hline %inserting double-line 
&\multicolumn{2}{c}{star-forming} & \multicolumn{2}{c}{passive galaxies} \\ [0.5ex] 
& A(amplitude) & $\alpha$ (slope) & A(amplitude) & $\alpha$ (slope) \\ [0.5ex] 
\hline  % inserts single-line 
$0.4<W^{\rm MgII}_{0}<0.8\,\rm \AA$ & $0.06\pm0.02$ &  & $0.06\pm0.01$ & \\[1ex] % Entering row contents 
$0.8<W^{\rm MgII}_{0}<1.5\,\rm \AA$ & $0.05\pm0.01$ & $-1.3 \pm 0.5$ & $0.04\pm0.01$ & $-1.4 \pm 0.3$\\[1ex]
$W^{\rm MgII}_{0}>1.5 \rm\,\AA$ & $0.02\pm0.01$ &  & $0.01\pm0.004$ & \\[1ex]
\hline 
$W^{\rm MgII}_{0}>1.0 \rm\,\AA$ & $0.05\pm0.01$ & $-1.2 \pm 0.4$ & $0.03\pm0.01$ & $-1.1 \pm 0.2$ \\[1ex] 
% [1ex] adds vertical space 
\hline % inserts single-line 
\end{tabular} 
\label{tab:covering_fraction} 
\end{table*} 
%
%% ==========================================

\subsubsection{\mgii\ covering fraction}
\label{covering}
%Because of the symmetry of finding galaxies around absorbers and finding absorbers around galaxies, the covering fraction can also be estimated from the ratio between the expected , , and , , which can be written as
%In the following, we will estimate the covering fraction by Equation~\ref{coveringfraction}. 
%where the denominator makes only use of lines-of-sight towards quasars to keep the same sampling function as that used in the numerator. As a result this quantity is not sensitive to the completeness correction discussed in Appendix A. 

To fully characterize the galaxy-absorber connection one should investigate both the incidence rate of galaxies around absorbers and the incidence rate of absorbers around galaxies, i.e. the absorption covering fraction $f_c$. This quantity is defined as
\begin{equation}
f_{c} \equiv N_{\rm abs}^{\rm [gal]}/N_{\rm QSO}^{\rm [gal]}\;\mathrm{,}
\end{equation}
where $N_{\rm QSO}^{\rm [gal]}$ is the total number of quasar sightlines around a given galaxy sample and $N_{\rm abs}^{\rm [gal]}$ is the total number of \mgii\ absorbers detected from those quasar sightlines. This quantity can also be estimated by counting galaxies around absorbers and quasars:
\begin{equation}
f_{c} = N^{\rm [abs]}_{\rm gal}/N^{\rm [QSO]}_{\rm gal}\;,
\label{coveringfraction}
\end{equation}
where $N_{\rm gal}^{\rm [abs]}$ (as estimated in Eq.~\ref{eq:Ngalabs}) is the total number of galaxies associated with \mgii\ absorbers and $N_{\rm gal}^{\rm [QSO]}$ is the total number of galaxies expected in that redshift interval. We will use this approach to estimate the gas covering fraction.
To do so we first calculate $N_{\rm gal}^{\rm [abs]}$, the expected total number of galaxies associated with a set of $N_{\rm abs}$ absorbers, spanning a redshift interval $\Delta z \ (0.4<z<0.6)$. It can be written as 
\begin{equation}
N_{\rm gal}^{\rm [abs]}= \langle N_{\rm gal}^{\rm abs}\rangle \times N_{\rm abs}^{\rm expected}\;,
\label{eq:Ngalabs}
\end{equation}
where $\langle N_{\rm gal}^{\rm abs}\rangle$ is the mean number of galaxies per absorber defined in Equation~\ref{main} and $N_{\rm abs}^{\rm expected}$ is the number of expected absorbers over the selected redshift range which is given by
\begin{equation}
N_{\rm abs}^{\rm expected} = dN/dz(W_{0}^{\rm MgII})\times \Delta z \times N_{\rm QSO}\;,
\end{equation}
with $dN/dz(W_{0}^{\rm MgII})$ being the \mgii\ incidence rate with a given $W_{0}^{\rm MgII}$ from \citet{ZhuMgIIcatalog}, $\Delta z$ being the redshift interval from 0.4 to 0.6, and $N_{\rm QSO}$ being the total number of quasars. In this analysis we only use quasars with $z>1$ to avoid possible contributions from galaxies physically clustered with the quasars. To estimate the denominator of Eq.~\ref{coveringfraction}, i.e., the expected number of galaxies falling in the selected redshift range $\Delta z$, we use galaxy counts around the selected quasars which allow us to estimate the total number of galaxies  $N_{\rm gal(all \ z)}^{\rm [QSO]}$ originating from all redshifts. We then quantify the number of galaxies expected within $\Delta z$ by 
\begin{equation}
N_{\rm gal}^{\rm [QSO]} = N_{\rm gal(\rm all \ z)}^{\rm [QSO]}\times \eta(\Delta z)
\label{Ngalaxy}
\end{equation}
where $\eta(\Delta z)$ is the fraction of galaxies with $0.4<z<0.6$ with respect to galaxies at all redshifts. To estimate $\eta(\Delta z)$, we select PRIMUS galaxies with robust redshift measurements and make use of the {\small WEIGHT} parameter in the catalog which accounts for the incompleteness due to target selection, fiber collision, and redshift quality (see Equation 1 in \citealt{MoustakasPRIMUS}). To alleviate the incompleteness of SDSS at faint end, we restrict this part of analysis to the sample of bright galaxies, i.e. selected with $i<20.6$. The estimated $\eta(\Delta z)$ are 0.18 and 0.49 for bright star-forming and passive galaxies, respectively.

% --------------------------------
\begin{figure}
\begin{center}
\includegraphics[scale=0.4]{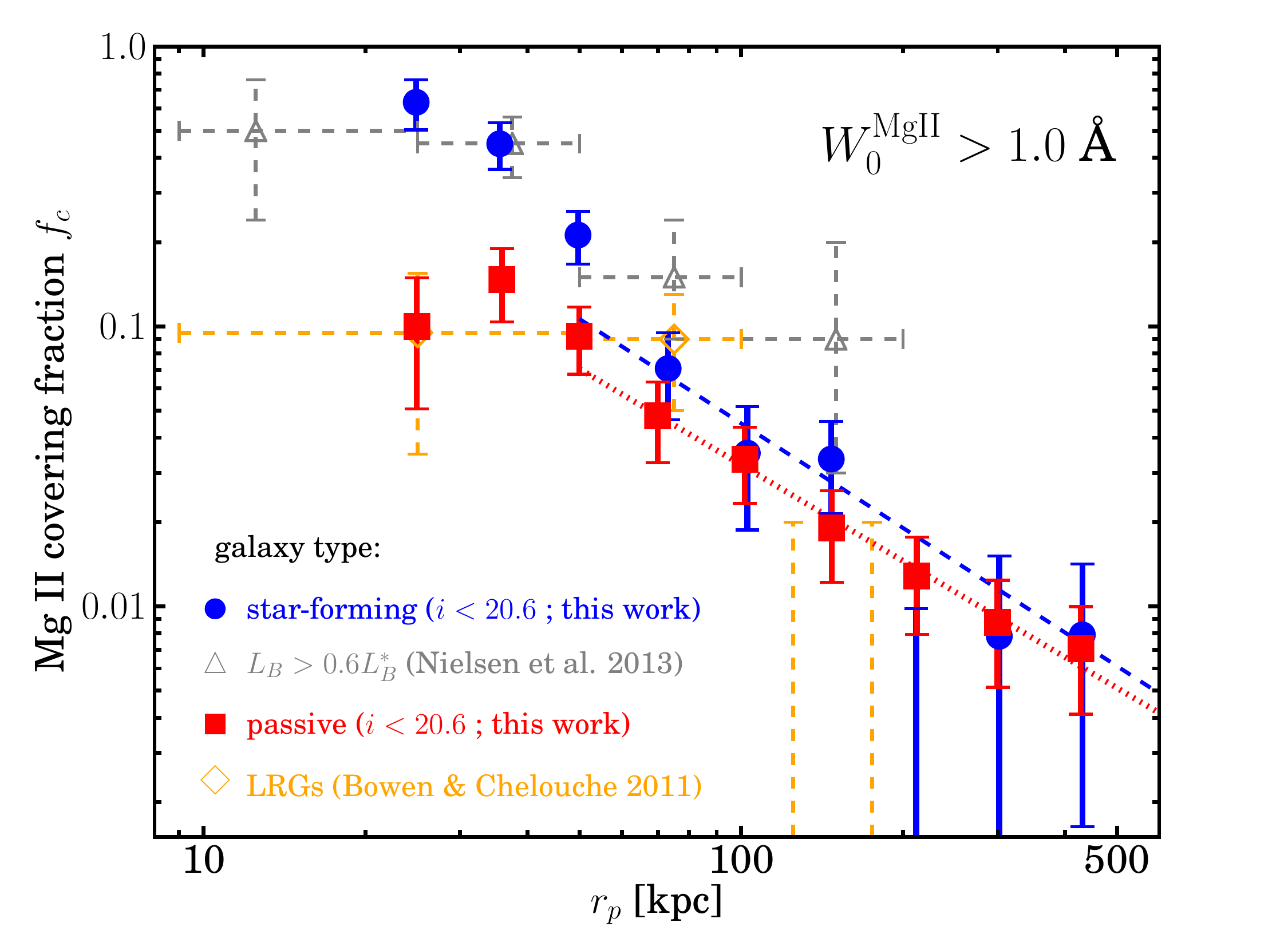}
\caption{The \mgii\ covering fraction around galaxies compared to the previous studies.
The grey triangles are the covering fraction of $L_{B}>0.6L_{B}^{*}$ galaxy sample from \citet{NielsenMgIIcatalogII}. 
The covering fraction of star-forming galaxies is consistent with their measurement (although they did not observe any color dependence of covering fraction). The orange diamonds are the covering fraction of luminous red galaxies (LRGs) from \citet{BowenLRG}. Considering similar colors, our bright passive galaxies have consistent covering fraction as \citet{BowenLRG}. The blue dashed line and red dotted line are the best-fit power laws beyond 50\,kpc.}
\label{fig:covering_fraction_cumulative}
\end{center}
\end{figure}

% - - - - - - - - - - - - - - - - - - - - - - - - - - - - - - - -
% figure
%---------------------------
\begin{figure*}[t]
\begin{center}
\includegraphics[scale=0.45]{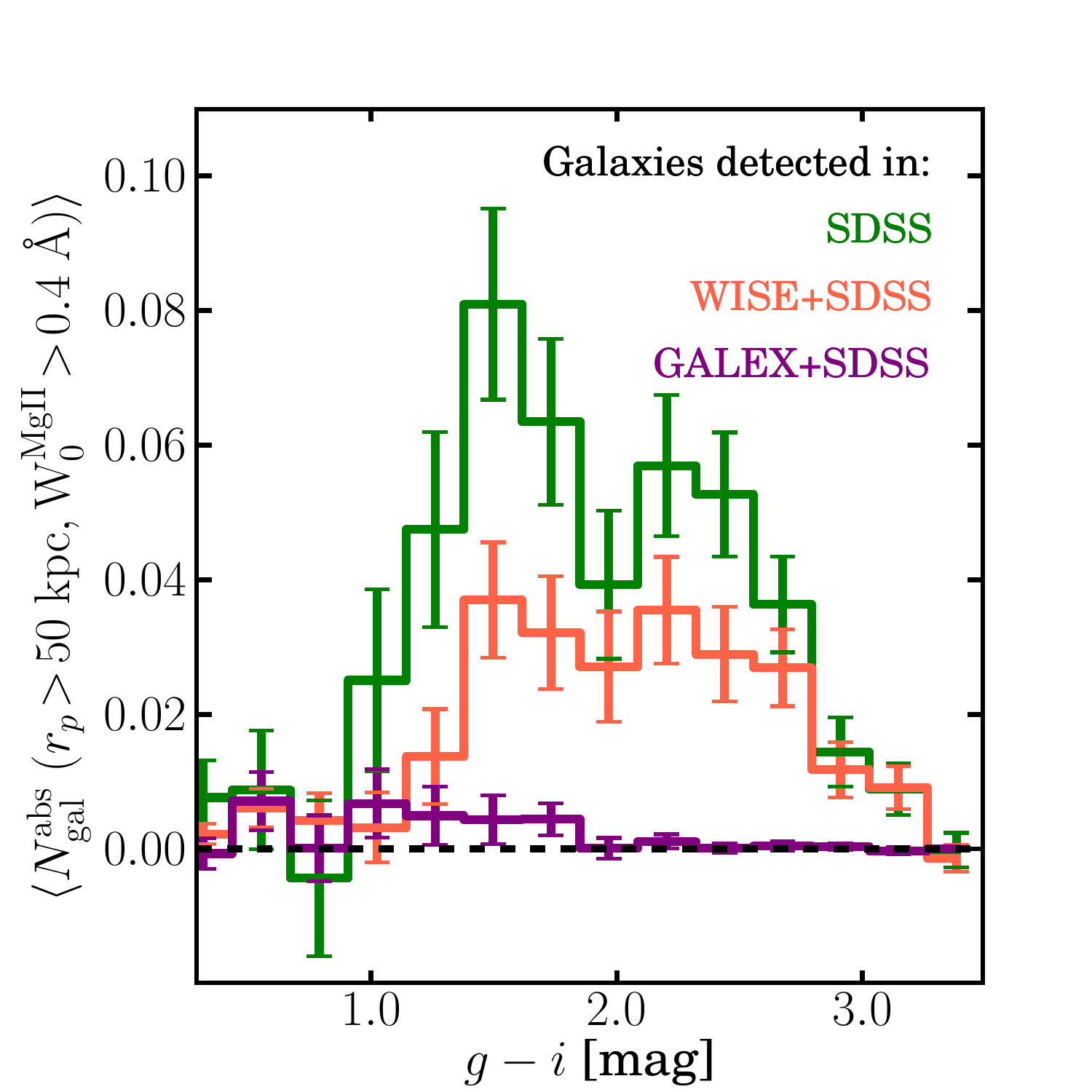}
\includegraphics[scale=0.45]{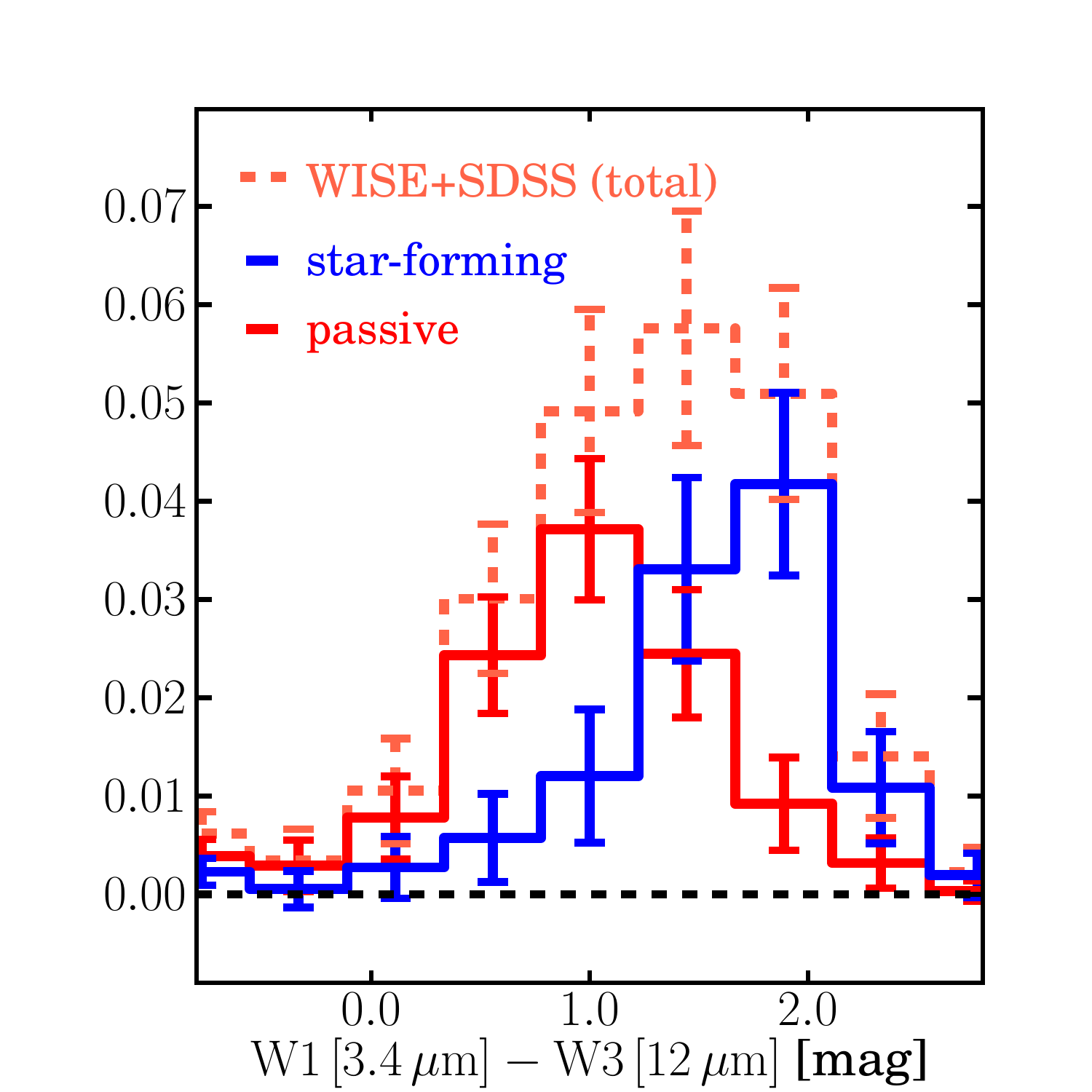}
\caption{\emph{Left:} The g-i color distributions of associated galaxies detected in SDSS, WISE and GALEX. SDSS, WISE and GALEX select different types of galaxies because of their wavelength coverages and sensitivities. With the highest sensitivity in W1 [3.4\,$\rm \mu m$] band, WISE tends to detect massive galaxies with relatively red galaxy color (in g-i). In contrast, GALEX selects young star forming galaxies bright in NUV with relatively blue galaxy color (in g-i). \emph{Right:} The W1-W3 color distribution of galaxies associated with \mgii\ absorbers. Orange dashed line: the associated galaxies detected in WISE and SDSS with $W_{0}^{\rm Mg\,II}>0.4$\,\AA \ from 50\,kpc to 200\,kpc. Blue: optical selected blue galaxies ($0.5<g-i<2.0$). Red: optical selected red galaxies ($2.0<g-i<3.0$). In general, optical-selected star-forming galaxies have more flux in W3 than W1 compared to passive galaxies due to strong $8\,\micron$ PAH emission.}
\label{fig:wise_color}
\end{center}
\end{figure*}
%-----------------------------------------------

Figure~\ref{fig:covering_fraction_binning} shows, for three differential bins of $W_{0}^{\rm MgII}$, the \mgii\ covering fraction as a function of impact parameter for both star-forming and passive galaxies with $i<20.6$. The average stellar masses of galaxies corresponding to the magnitude cut are $\langle \log M_{\ast}/{\rm M_{\odot}}\rangle\sim10.6$ (star-forming) and $\sim10.9$ (passive). We find that the covering fraction of strong absorbers around star-forming galaxies is more than 2 times higher than around passive galaxies at impact parameters smaller than about 50\,kpc, consistent with what was found in Section~\ref{galaxy_type_mgii}. On the other hand, beyond 50\,kpc, there is no significant difference between the \mgii\ covering fraction of two types of galaxies and the covering fraction can be described by a power-law functional form 
\begin{equation}
f_c= {\rm A}\,\left(\frac{r_{p}}{100\,{\rm kpc}}\right) ^{\alpha} \ {\rm for} \ r_{p}>50\,{\rm kpc}.
\end{equation}
At impact parameters greater than 50\,kpc, we find similar spatial dependences for different bins of \mgii\ absorption strength. To characterize them we fit the covering fraction with a single slope for each type of galaxies and report the best-fit parameters shown in Table~\ref{tab:covering_fraction}. 
On large scales, the \mgii\ covering fraction appears to be consistent with the scale dependence of galaxy-galaxy cross-correlation function, suggesting that \mgii\ absorbers around galaxies at large scales may trace the overall mass distribution. Note that by extrapolating the power-law functions down to 20\,kpc, the functions can still reproduce the covering fraction around passive galaxies while they will underestimate the covering fraction of strong absorbers around star-forming galaxies. 

%as shown in \citet{2014MNRAS.tmp..377Z}. 

Figure~\ref{fig:covering_fraction_cumulative} shows comparisons between our measurements and previous studies for $W^{\rm Mg\,II}_{0}> 1.0$\,\AA. In \citet{BowenLRG}, the authors used photo-z of LRGs to find galaxy-\mgii\ absorber pairs and estimated the covering fraction of \mgii\ absorbers around LRGs. With similar color of their LRGs, our passive galaxies have consistent \mgii\ covering fraction. We also compare our results to the covering fraction of high-luminosity galaxies in \citet{NielsenMgIIcatalogII}. Our bright star-forming galaxies have a consistent covering fraction with their measurement (although they did not observe any dependence on galaxy color). The best-fit parameters of the corresponding dependence are given in Table~\ref{tab:covering_fraction}.
%% -------------------
% figure
%% -----------------------------
\begin{figure}[th]
\begin{center}
\includegraphics[scale=0.45]{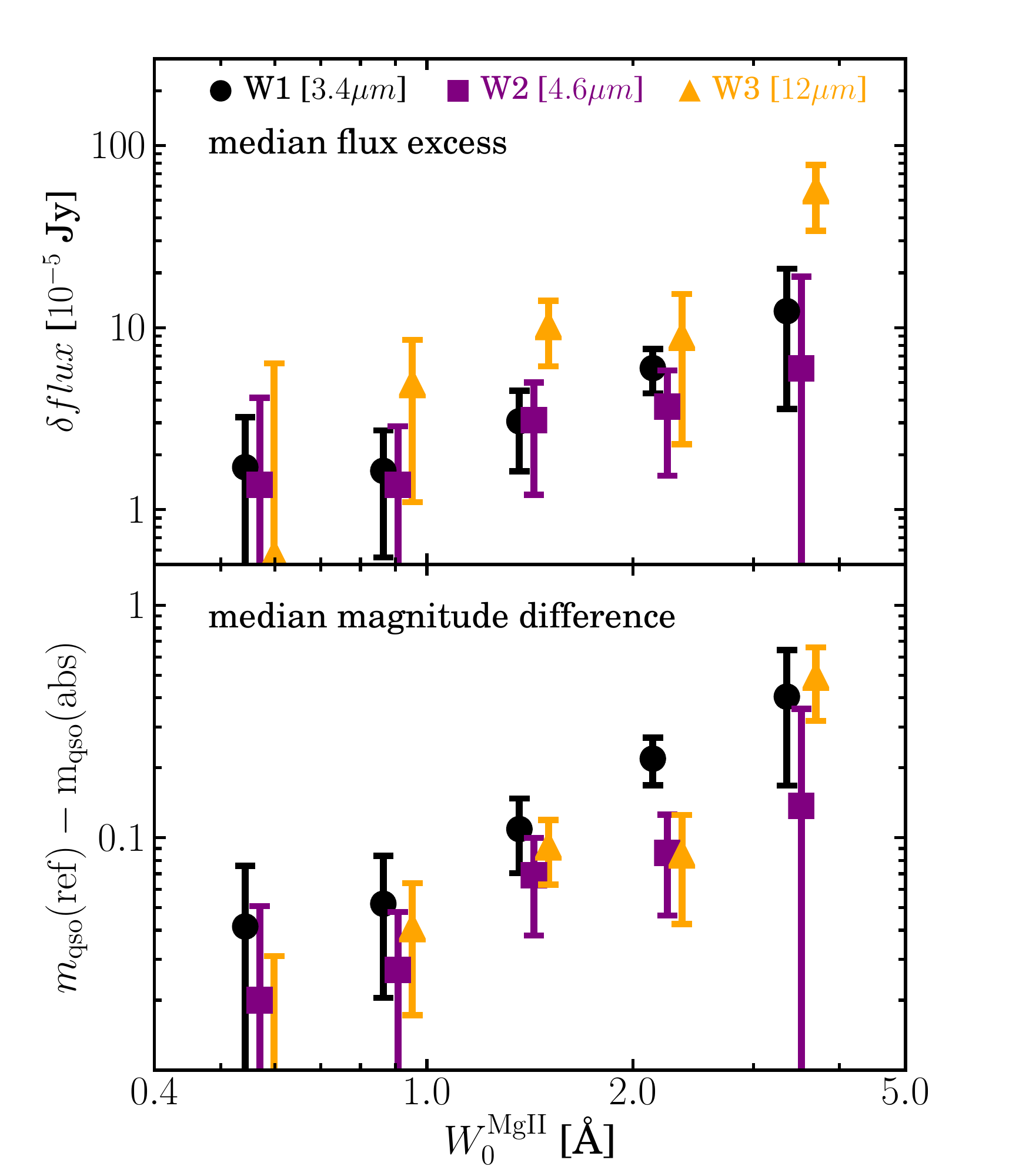}
\caption{Quasar brightening effect due to the presence of a galaxy associated with a \mgii\ absorber. \emph{Top:} median flux excess associated with absorbers as a function of $W_{0}^{\rm Mg\,II}$. The spectral energy distribution of the excess is consistent with that of a star forming galaxy with strong $8\,\micron$ PAH emission.\emph{Bottom:} median quasar magnitude shift due to the presence of an absorber. The amplitude of this effect can be as large as 0.5 magnitude for the strongest absorbers.}
\label{fig:wise_quasar}
\end{center}
\end{figure}
%---------------------------

%% ========================================
\subsection{Correlations with WISE and GALEX}
\label{sec:GALEX_WISE}

We now complement our analysis with photometric information from the WISE and GALEX datasets. The angular resolution of WISE [$3.4\,{\rm \mu m}$] band and GALEX NUV band is about $6\arcsec$, which is 4 times that of SDSS. This restricts our number count analysis to impact parameters greater than about 50\,kpc. On smaller scales, galaxies falling within the angular resolution of quasars will affect its apparent brightness by providing an extra flux contribution and/or by dimming the quasar light due to dust extinction.

%% ========================================
\subsubsection{Large scale correlations}

We first match sources detected in GALEX and/or WISE with galaxies detected in SDSS within a radius of $6\arcsec$. We find that the vast majority of WISE and/or GALEX galaxies associated with \mgii\ absorbers are also detected in SDSS. The left panel of Figure~\ref{fig:wise_color} shows the $g-i$ color distribution of galaxies found in excess beyond 50 kpc of \mgii\ absorbers with $W_{0}^{\rm Mg\,II}>0.4$\,\AA. The green, orange, and purple curves show sources detected in SDSS, WISE+SDSS and GALEX+SDSS, respectively.

All the sources detected in GALEX appear to be blue in $g-i$. They are on average bluer than the blue population of galaxies selected from SDSS only. This reflects that GALEX traces UV-bright star-forming galaxies at those redshifts.
We find the average NUV magnitude to be about 21.5. 
According to \citet{Kennicutt1998}, this corresponds to an average SFR $\sim 10 \ \rm M_{\odot}\,yr^{-1}$.

The $g-i$ color distribution of sources selected in both SDSS and WISE shows a relatively higher sensitivity to red galaxies. To better understand the relation between optical and infrared colors we select star-forming and passive galaxies from their $g-i$ colors (using the same selection cuts asin Equation~\ref{eq:colorcut}) and investigate their infrared color distributions. To do so we have estimated the WISE color W1 [$3.4\,{\rm\mu m}$]-W3 [$12\, {\rm\mu m}$]. The result is shown in the right panel of Figure~\ref{fig:wise_color}. The orange dashed line shows all WISE+SDSS galaxies distribution, blue data points represent optical-selected star-forming galaxies, and red data points represent optical-selected passive galaxies. This figure shows that, at $z\sim0.5$ optically star-forming galaxies have red W1-W3 colors. This is because star-forming galaxies produce strong $8\,{\rm\mu m}$ polycyclic aromatic hydrocarbon (PAH) emission lines visible in the W3 band \citep[e.g.][]{SmithPAH} at those redshifts. This analysis shows that the types of galaxy associated with absorbers found with the SDSS survey are consistent with those found at infrared wavelengths with the WISE survey.

%% ===========================================
\subsubsection{Small scale correlations}

On scales smaller than about 50\,kpc, galaxies associated with absorbers fall within the angular resolution of the WISE quasar and affect its apparent brightness. In this analysis, we match our SDSS quasar sample with WISE full-sky catalog with a $6\arcsec$ radius. %\citep{YanWISE}
We use WISE standard aperture magnitudes (which has a size of $8\arcsec$) which contain all flux including quasars and galaxies on top of them. We limit our sample with absorber redshift from 0.4 to 0.6, calculate the median WISE flux of quasars with \mgii\ absorbers as a function of \mgii\ absorption strength, and compare them to the median WISE flux of reference quasars. The error is estimated by bootstrapping. Note that quasars with \mgii\ absorbers and their reference quasars have similar redshift and i-band magnitude distributions. 

The top panel of Figure~\ref{fig:wise_quasar} shows the flux differences in WISE 1, 2, and 3 bands as a function of $W_{0}^{\rm Mg\,II}$. We find that quasars with \mgii\ absorbers are systematically brighter than reference quasars. The flux difference increases for stronger absorbers. This is in part because stronger absorbers tend to be closer to star-forming galaxies. We note that the excess flux in W3 [$12\,{\rm \mu m}$] tends to be higher than that of W1 [$3.4\,{\rm\mu m}$]. This is consistent with the $8\, {\rm \mu m}$ PAH contribution discussed above.
 
The bottom panel of the figure shows how the IR flux contribution from the absorbing galaxies affects quasar magnitudes. The median magnitude shift induced by the presence of a \mgii\ absorber, $\delta m=\rm m_{QSO}(\rm ref)-m_{QSO}(\rm abs)$, is shown as a function of $W_0^{\rm Mg\,II}$. On average, the median quasar brightness changes by about 0.1 and reaches 0.5 magnitude for the strongest absorbers. This result indicates that selecting quasars with WISE magnitudes and colors could potentially result in a selection bias for quasars with the presence of \mgii\ absorbers. Future surveys, for example SDSS-IV\footnote{\url{http://www.sdss3.org/future/}}, that use WISE magnitudes and colors to select quasars should take into account such bias. In the UV, the presence of an intervening absorber gives rise to extinction effects. This was recently studied by \citet{Menarddust}.

%% ===================================================
\section{Discussion}

We have shown that cool gas traced by \mgii\ absorption is found around both star-forming galaxies and passive galaxies, with a similar incidence rate at impact parameters greater than about 50 kpc. In contrast, at smaller impact parameters we find that the strength of \mgii\ absorption depends on the level of star formation of the central galaxy. Our results are consistent with other types of observational constraints on the galaxy-absorber connection: \citet{Zibetti2007} used stacked SDSS images to measure the mean flux excess correlated with the presence of \mgii\ absorbers and found stronger \mgii\ absorbers to be preferentially associated with bluer emission. \citet{BordoloiMgII} used stacked galaxy spectra to measure the mean \mgii\ absorption induced by the presence of galaxies along the line-of-sight and showed that the mean \mgii\ absorption strength around star-forming galaxies is higher than around passive galaxies. 
Our analysis has, in addition, allowed us to show that while \mgii\ absorption is commonly found around passive galaxies, no correlation between equivalent width and galaxy properties can be detected. This is in contrast to the relation observed between star-forming galaxies and absorbers where the mean equivalent width depends on the color and/or star formation rate of the galaxies. The basic dichotomy between star-forming and passive galaxies is therefore reflected in the properties of the cool gas in the CGM. We also find that strong absorbers tend to be found along the minor axis of star-forming galaxies at small impact parameters, indicating that some of the gas detected in absorption is likely associated with outflows from galaxies. This is consistent with what was reported by \citet{BordoloiMgII}, \citet{BoucheMgII} and \citet{KacprzakAZ}, and is similar to the measurements of \caii\ absorption by \citet{ZhuCaII}. Our results also show that at impact parameters greater than 50\,kpc, the covering fraction of strong \mgii\ absorbers declines with scale proportionally to the overall mass distribution for both star-forming and passive galaxies. Here we note that \citet{ZhuMgII} found a similar result by measuring the \emph{total} \mgii\ absorption around SDSS luminous red galaxies. By measuring the mean absorption correlated with the presence of such galaxies, these authors showed that the average \mgii-to-dark matter ratio is roughly scale-independent. Strong \mgii\ absorbers therefore trace the overall \mgii\ absorption field on large scales.

\subsection{The connection to baryons in galaxy halos}

It is interesting to point out the contrast between the ubiquitous presence of cool, low-ionized gas traced by \mgii\ around both star-forming and passive galaxies and the distribution of highly ionized gas traced by \ovi\ absorption which currently is only detected in the halo of star-forming galaxies \citep{TumlinsonOVI}. However, despite such a difference, the two gas distributions share a common property: excess absorption is detected when the galaxy specific star formation rate $sSFR\gtrsim 10^{-11}\,{\rm yr}^{-1}$. This threshold is found to be the same for both phases. It appears to be a characteristic sSFR value for determining the gaseous properties of the CGM.

Similarly, we detect excess absorption when the galaxy star formation rate ${\rm SFR}\gtrsim 1\,\rm M_\odot/yr$ and note that this threshold is similar to the one for which blueshifted \mgii\ self-absorption is seen in galaxies \citep[e.g.][]{Weiner2009,Rubin2013, Bordoloioutflow}. The interpretation of such studies is usually limited by the lack of information on the spatial scales over which the gas flow is occurring. The similarity of our results suggests that the two lines of investigation are probing the same material. If so, the gas seen as blueshifted self-absorption could extend up to scales greater than a few kpc, as typically assumed by these authors when inferring mass outflow rates from such measurements.

Having characterized the covering fraction of \mgii\ absorbers as a function of impact parameter, we can attempt to estimate the typical amount of cool HI gas traced by \mgii\ absorbers residing in the CGM of $z\sim0.5$ galaxies. To do so we can use the empirical relation between \mgii\ equivalent width and median HI column density measured by \citet{2006ApJ...636..610R} and quantified by \citet{2009MNRAS.393..808M}: $\langle N_{\rm HI} \rangle(W_0^{\rm Mg\,II})$ (their Equation 5). Using this relation we can write
\begin{eqnarray}
M_{\rm HI}^{\rm CGM}(>W_0^{\rm Mg\,II}) &\sim& 2\pi \int_{20\,{\rm kpc}}^{150\,{\rm kpc}} \langle N_{\rm HI} \rangle f_{c}(r_{p})\,r_{p}\,{\rm d}r_{p}\, \nonumber\\
\,
\end{eqnarray}
where $f_{c}(r_{p})$ is the \mgii\ covering fraction and $r_{p}$ is the projected distance. The inner impact parameter limit (20 kpc) is selected based on the range for which we have robust covering fraction measurements, while the outer impact parameter limit (150 kpc) corresponds to the maximum impact parameter of galaxy-absorber pairs for which COS-Halos team searched.
To estimate this quantity numerically we consider the cool HI gas traced by $W^{\rm MgII}_{0}> 1\rm\,\AA$ ($\langle W^{\rm Mg\,II}_{0}\rangle\simeq1.6 \rm\, \AA$), corresponding to the covering fraction shown in Figure~\ref{fig:covering_fraction_cumulative}. For star-forming galaxies with $\langle \log M_\ast/M_\odot \rangle\sim 10.6$ we find
\begin{eqnarray}
\log M_{\rm HI}^{\rm CGM}/{\rm M_\odot} &\sim& {9.5}.
\end{eqnarray}
For passive galaxies with $\langle \log M_\ast/M_\odot \rangle\sim 10.9$ we find
\begin{eqnarray}
\log M_{\rm HI}^{\rm CGM}/{\rm M_\odot} &\sim& {9.2}.
\end{eqnarray}
The above numbers correspond only to HI gas traced by \mgii\ absorbers stronger than $1\,{\rm \AA}$. The estimate can be considered as a lower limit for the total amount of HI found in galaxy halos.
These values imply that the ratio $M_{\rm HI}^{\rm CGM}/M_\ast$ is about four times lower around passive galaxies than around star-forming ones. 

Recently, \citet{Werk2014} used HST-COS observations to probe the gaseous distribution around $z\sim 0.2$ $L\sim L_\ast$ galaxies. Using photo-ionization models they inferred that such galaxies are surrounded by cool ($T \sim 10^4\,$K) gas amounting to at least $\log M_{\rm H}/M_\odot = 10.4$ within about 150\,kpc. Most (99\%) of this gas is found to be ionized, implying that the neutral component probed by the lines-of-sight of the COS-Halos survey (excluding a few damped systems) amounts to $\langle \log M_{\rm HI}/M_\odot \rangle\sim 8.4$. This is a factor 5-10 lower than the amount of HI probed by \mgii\ absorbers presented above. We note that given the low covering fraction for strong \mgii\ absorbers on such scales: $f_c(100\,{\rm kpc})\sim 0.05$, most of randomly-selected lines-of-sight are not expected to intercept such clouds. Interestingly, the above numbers indicate that most of the neutral hydrogen probed by metal absorbers is located in strong \mgii\ absorbers, despite their low covering fraction. 
An additional neutral gas contribution can be associated with pristine or low metallicity gas. Such clouds would not give rise to strong \mgii\ absorbers. 

We also note that the spatial dependence of the covering fraction of strong \mgii\ absorbers (derived in Section~\ref{covering}) can be converted into a minimum value for the \emph{mean} absorption equivalent width as a function of impact parameter. By selecting absorbers with $W_0^{\rm Mg\,II}>1.0$\,\AA\ with $1.6$\,\AA\ mean absorption and focusing on red galaxies, we find that 
\begin{equation}
\langle W_0^{\rm Mg\,II} \rangle(r_p) \sim 0.05\,\left( \frac{r_p}{100\,{\rm kpc}} \right)^{-1.1} {\rm \AA} \;.
\end{equation}
This dependence can be compared to the results obtained by \citet{ZhuMgII}. These authors measured the mean absorption around galaxies averaged over all lines-of-sight. By comparing the two sets of results, we find that absorbers with $W_0^{\rm Mg\,II}>1\,$\AA\ contribute to about half of the total absorption signal. By selecting  systems with $W_0^{\rm Mg\,II}>0.4\,$\AA\ we find this contribution to be comparable to the mean absorption level. This indicates that, on large scales around galaxies, strong \mgii\ absorbers dominate the total \mgii\ absorption budget.

Our results also allow us to reveal that the covering fraction for cool gas around galaxies appears to change as a function of redshift. At $z\sim 0.5$ we found $f_c \propto r_p^{-1}$ for \mgii\ absorbers at impact parameters greater than about 50\,kpc around both star-forming and passive galaxies. As mentioned above, this relation steepens on smaller scales around star-forming galaxies. Using a statistical analysis, \cite{2010ApJ...717..289S} inferred the radial dependence of the covering fraction of cool gas around star-forming (Lyman break) galaxies at $2\lesssim z \lesssim 3$. They derived a significantly shallower radial dependence: $f_c \propto r_p^{-\gamma}$ with $0.2\lesssim \gamma \lesssim 0.6$, depending on the transition. The comparison of the two analyses therefore shows that the radial dependence of metals in cool gas appears to steepen from $z\simeq2-3$ down to $z\sim0.5$.

Finally, if we assume that the neutral gas traced by \mgii\ absorbers is not only enriched in metals in the gas phase but also dusty, it implies a CGM dust mass of $M_{\rm dust}^{\rm CGM}\sim 4\times 10^{7}\,{\rm M_\odot}$ {using the global dust-to-gas ratio in \mgii\ clouds \citep[$\sim1/50$;][]{Menarddust}. This value is consistent with the findings of \citet{2010MNRAS.405.1025M} and \citet{Peek2014} who statistically mapped out the distribution of dust in galaxy halos using reddening measurements and inferred its total mass. It also implies that most of the circum-galactic dust is associated with \mgii\ absorbers.

\section{Summary}

We cross-correlate about 50,000 \mgii\ absorbers with photometric sources from SDSS, WISE, and GALEX to study the properties of cool (T$\sim10^4\,$K) gas in the circum-galactic medium. Using the SDSS survey and focusing on the redshift range $0.4<z<0.6$ we statistically extracted about 2,000 galaxy-absorber pairs which have allowed us to explore the relationship between absorber and galaxy properties with an unprecedented sensitivity. Our results are summarized as follows:
\begin{itemize}

\item \mgii\ absorbers are associated with both star-forming and passive galaxies, with a comparable incidence rate at impact parameters greater than 50\,kpc. However each galaxy type exhibits a different behavior: within 50\,kpc \mgii\ equivalent width correlates with the presence of star-forming galaxies but not with that of passive galaxies.

\item The correlation between the presence of cool gas traced by strong \mgii\ absorbers and edge-on star-forming galaxies is stronger along the minor axis of galaxies at impact paramters reaching about 50\,kpc, suggesting that some of the gas is associated with outflows, consistent with previous studies. In contrast, we find \mgii\ absorbers to be isotropically distributed around edge-on passive galaxies.

\item We measure the average excess \mgii\ equivalent width $\langle\delta W_{0}^{\rm Mg\,II}\rangle$ as a function of galaxy properties and find $\langle\delta W_{0}^{\rm Mg\,II}\rangle\propto SFR^{1.2}$, $sSFR^{0.5}$ and $M_\ast^{0.4}$ for star-forming galaxies. These observational results can be used to constrain models of galaxy formation and feedback processes.

\item We characterize the covering fraction of \mgii\ absorption as a function of impact parameter and find that it is about 2-10 times higher for star-forming galaxies than passive ones on scales smaller than about 50\,kpc. The covering fractions appear to be comparable on scales larger than 50\,kpc and follow the spatial dependence of the galaxy correlation function. 

\item Using GALEX and WISE we show that the UV and IR properties of the galaxies correlated with absorbers are consistent with the dependence found with optical data. On scales smaller than 50\,kpc, we show how the presence of $z\sim0.5$ \mgii\ absorbers modifies the brightness and colors of quasars, consistent with dust reddening at short wavelengths and excess emission at IR wavelengths. We also show how the selection of quasars in the WISE magnitude and color space can be biased by the presence of $z\sim0.5$ galaxies detectable through their optical metal absorption. Future surveys using WISE information to select quasars need to take into account this bias. 

\end{itemize}
This work shows that the dichotomy between star-forming and passive galaxies is reflected in the CGM traced by low-ionized gas. This trend is similar to what has been reported for the distribution of higher-ionization species such as \ovi\ \citep{TumlinsonOVI}.\\

This statistical analysis demonstrates that galaxy-absorber correlations can be measured without prior knowledge on galaxy redshifts. This has allowed us to study the distribution of cool gas around galaxies and its dependence on galaxy properties. Our results present a set of constraints for theoretical models of galaxy growth and feedback. The analysis can be pushed toward higher redshifts as deeper imaging over large areas becomes available, such as the ongoing Hyper-Suprime Cam Survey\footnote{\url{http://www.naoj.org/Projects/HSC}} or the planned DECam Legacy Survey.

%%-------------------------
% figure
%% ----------------------- 
\begin{figure*}
\begin{center}
\includegraphics[scale=0.45]{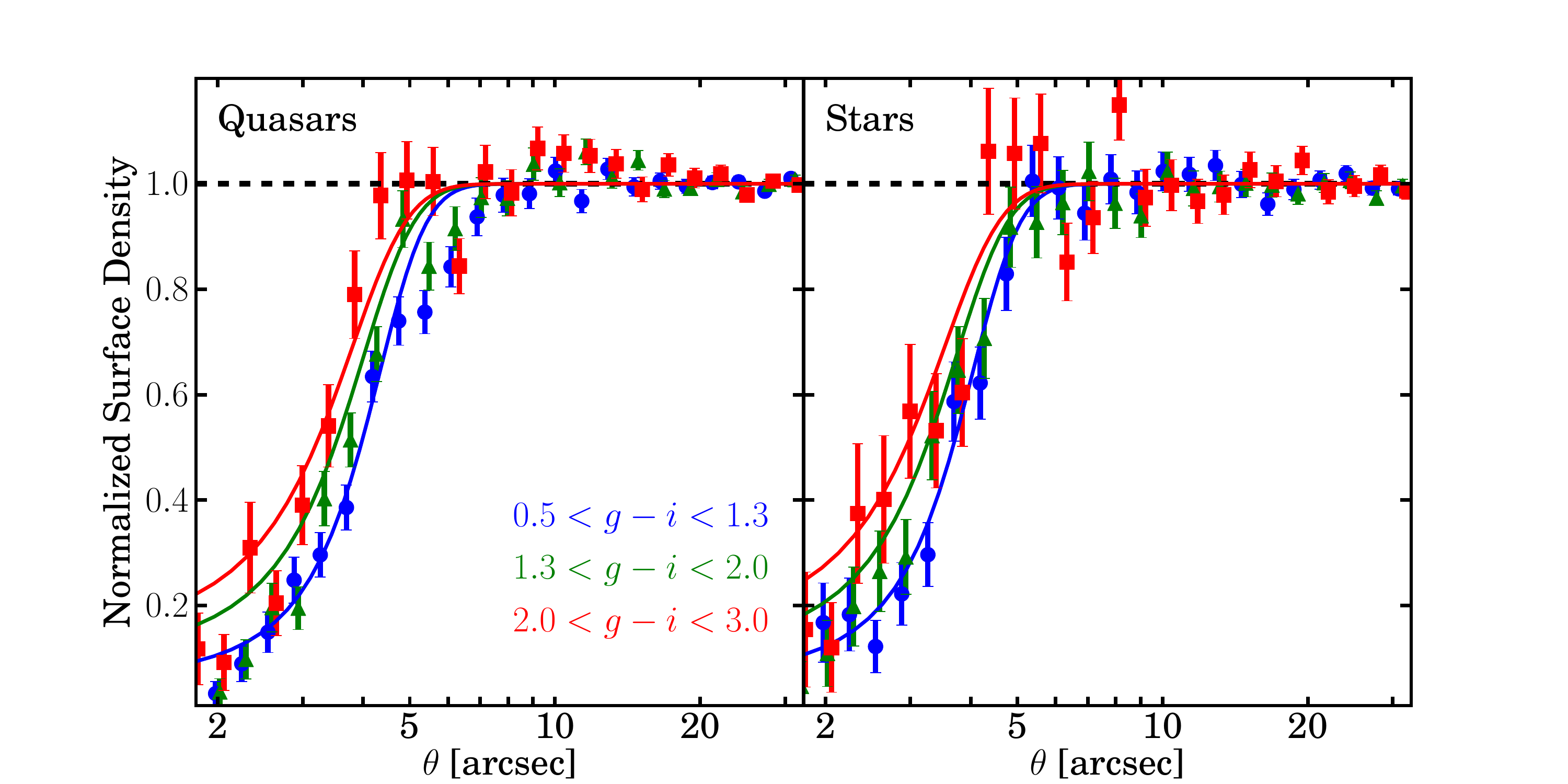}
\caption{The normalized surface density of galaxies around quasars and stars. \emph{Left}: quasars. \emph{Right}: stars. We estimate the surface densities of galaxies around quasars and stars and normalize them with the average surface densities at large scale (from $85\arcsec$ to $185\arcsec$). The shape can be well parameterized by the inverse of Eq.~\ref{correction} shown as the solid curves in the figure. We find that red galaxies have lesser blending effect than blue galaxies. It could be due the difference of physical properties of galaxies; nevertheless, the detailed understand is beyond the scope of this paper.}
\label{fig:Appendix_correction}
\end{center}
\end{figure*}

\acknowledgements
This work was supported by NSF Grant AST-1109665, the Alfred P. Sloan foundation and a grant from Theodore Dunham, Jr., Grant of Fund for Astrophysical Research. TWL was supported by William Gardner Fellowship.  We thank the PRIMUS team and John Moustakas for providing us with the star formation rate and stellar mass measurements of the PRIMUS sample. We thank Jason Tumlinson for useful discussions.  

Funding for the SDSS and SDSS-II has been provided by the Alfred P. Sloan Foundation, the Participating Institutions, the National Science Foundation, the U.S. Department of Energy, the National Aeronautics and Space Administration, the Japanese Monbukagakusho, the Max Planck Society, and the Higher Education Funding Council for England. The SDSS Web Site is http://www.sdss.org/. Funding for SDSS-III has been provided by the Alfred P. Sloan Foundation, the Participating Institutions, the National Science Foundation, and the U.S. Department of Energy Office of Science. The SDSS-III web site is http://www.sdss3.org/.

SDSS-III is managed by the Astrophysical Research Consortium for the Participating Institutions of the SDSS-III Collaboration including the University of Arizona, the Brazilian Participation Group, Brookhaven National Laboratory, University of Cambridge, Carnegie Mellon University, University of Florida, the French Participation Group, the German Participation Group, Harvard University, the Instituto de Astrofisica de Canarias, the Michigan State/Notre Dame/JINA Participation Group, Johns Hopkins University, Lawrence Berkeley National Laboratory, Max Planck Institute for Astrophysics, Max Planck Institute for Extraterrestrial Physics, New Mexico State University, New York University, Ohio State University, Pennsylvania State University, University of Portsmouth, Princeton University, the Spanish Participation Group, University of Tokyo, University of Utah, Vanderbilt University, University of Virginia, University of Washington, and Yale University.

This publication makes use of data products from the Wide-field Infrared Survey Explorer, which is a joint project of the University of California, Los Angeles, and the Jet Propulsion Laboratory/California Institute of Technology, funded by the National Aeronautics and Space Administration.

%----------------------------------------------------
%% =========================================
% Appendix
%% ========================================

\appendix 
\section{Correction for blending effect}
SDSS has relatively small average angular resolution ($1.4\arcsec$) which enables us to probe small impact parameter region. Nevertheless, the size of the angular resolution will affect the detectability of galaxies within certain scales ($<4\arcsec$) due to the blending with quasars. To test this effect and correct for it, we estimate the surface density of galaxies around quasars. We select quasars with redshift higher than 1 to avoid the physical clustering between quasars and galaxies, and with i-band magnitudes brighter than 19.8 to have a similar magnitude limit of quasars with detected \mgii\ absorbers. We search galaxies around selected quasars and calculate the surface density of galaxies with a given color bin. Then, we normalize the surface densities by average surface densities calculated from large scale, $85\arcsec$ to $185\arcsec$.

The left panel of Figure~\ref{fig:Appendix_correction} shows the normalized surface densities of galaxies with three color bins around selected quasars. We find that the effect of blending extends to about $4\arcsec$, corresponding to 30 kpc at redshift 0.5. Not only that, the normalized surface densities depend on galaxy colors; red galaxies have less effect than blue galaxies. This effect could be due to the properties of galaxies within different color bins, especially the magnitude distribution and size. However, the detailed understanding of this color dependence is beyond the scope of this paper.

To correct for this blending effect, we introduce an empirical correction function, $w$, with a single Gaussian as 
\begin{equation}
w(\theta,C)= 1+ \frac{A}{C}\times e^{\frac{-\theta^{2}}{2\sigma^{2}}} 
\label{correction}
\end{equation}
where C and $\theta$ are the $g-i$ color and angular distance between galaxies and quasars in arcsec. Given an angular distance and galaxy color, $w(\theta,C)$ represents the correct number count without the blending effect.
We fit the inverse of the correction function, $1/w$, to normalized surface densities and the best-fit parameters yield to $A=16.2\pm3.4$ and $\sigma=1.65\arcsec\pm0.07$. The solid curves in Figure~\ref{fig:Appendix_correction} are the best-fit normalized surface densities from our formula. The width of the Gaussian $\sigma\sim1.65\arcsec$ is comparable with the size of the average angular resolution, suggesting that Gaussian is a reasonable functional form to describe the blending effect.

To test the robustness of this measurement, we also randomly select about 37,500 stars from SDSS with i-band magnitude from 17 to 19.8 and apply the same analysis. The right panel of Figure~\ref{fig:Appendix_correction} shows the results from stars. The best-fit parameters from stars are consistent with quasars within one sigma. Therefore, we confirm that the behavior is mainly driven by blending and use our parameterized correction function (Equation~\ref{correction}) to correct the number count of galaxies around \mgii\ absorbers in Sections 3.1.1, 3.1.2, and 3.1.4.


\begin{thebibliography}{}


\bibitem[Abazajian et al.(2009)]{AbazajianDR7} Abazajian, K.~N., Adelman-McCarthy, J.~K., Ag{\"u}eros, M.~A., et al.\ 2009, \apjs, 182, 543 
\bibitem[Bergeron(1986)]{Bergeron1986} Bergeron, J.\ 1986, \aap, 155, L8
\bibitem[Bordoloi et al.(2011)]{BordoloiMgII} Bordoloi, R., Lilly, S.~J., Knobel, C., et al.\ 2011, \apj, 743, 10
\bibitem[Bordoloi et al.(2013)]{Bordoloioutflow} Bordoloi, R., Lilly, 
S.~J., Hardmeier, E., et al.\ 2013, arXiv:1307.6553 
\bibitem[Bouch{\'e} et al.(2012)]{BoucheMgII} Bouch{\'e}, N., Hohensee, W., Vargas, R., et al.\ 2012, \mnras, 426, 801 
\bibitem[Bowen \& Chelouche(2011)]{BowenLRG} Bowen, D.~V., \& Chelouche, D.\ 2011, \apj, 727, 47
\bibitem[Coil et al.(2011)]{CoilPRIMUS} Coil, A.~L., Blanton, M.~R., Burles, S.~M., et al.\ 2011, \apj, 741, 8
\bibitem[Cool et al.(2013)]{CoolPRIMUS} Cool, R.~J., Moustakas, J., Blanton, M.~R., et al.\ 2013, \apj, 767, 118

\bibitem[Kacprzak et al.(2011)]{Kacprzak2011} Kacprzak, G.~G., Churchill, C.~W., Evans, J.~L., Murphy, M.~T., 
\& Steidel, C.~C.\ 2011, \mnras, 416, 3118 
\bibitem[Kacprzak et al.(2012)]{KacprzakAZ} Kacprzak, G.~G., Churchill, C.~W., \& Nielsen, N.~M.\ 2012, \apjl, 760, L7 
\bibitem[Kennicutt(1998)]{Kennicutt1998} Kennicutt, R.~C., Jr.\ 1998, \araa, 36, 189 

%\bibitem[Lehner et al.(2013)]{2013ApJ...770..138L} Lehner, N., Howk, J.~C., Tripp, T.~M., et al.\ 2013, \apj, 770, 138 
\bibitem[Martin et al.(2005)]{MartinGALEX} Martin, D.~C., Fanson, J., Schiminovich, D., et al.\ 2005, \apjl, 619, L1
\bibitem[M{\'e}nard \& Chelouche(2009)]{2009MNRAS.393..808M} M{\'e}nard, B., \& Chelouche, D.\ 2009, \mnras, 393, 808 
\bibitem[M{\'e}nard et al.(2010)]{2010MNRAS.405.1025M} M{\'e}nard, B., Scranton, R., Fukugita, M., \& Richards, G.\ 2010, \mnras, 405, 1025 
\bibitem[M{\'e}nard et al.(2011)]{MenardOII} M{\'e}nard, B., Wild, V., Nestor, D., et al.\ 2011, \mnras, 417, 801 
\bibitem[M{\'e}nard \& Fukugita(2012)]{Menarddust} M{\'e}nard, B., \& Fukugita, M.\ 2012, \apj, 754, 116


\bibitem[Moustakas et al.(2013)]{MoustakasPRIMUS} Moustakas, J., Coil, A.~L., Aird, J., et al.\ 2013, \apj, 767, 50
\bibitem[Morrissey et al.(2007)]{GALEX_ref} Morrissey, P., Conrow, T., Barlow, T.~A., et al.\ 2007, \apjs, 173, 682
\bibitem[Mostek et al.(2012)]{Mostek2012} Mostek, N., Coil, A.~L., Moustakas, J., Salim, S., \& Weiner, B.~J.\ 2012, \apj, 746, 124 

\bibitem[Nielsen et al.(2013a)]{NielsenMgIIcatalogI} Nielsen, N.~M., Churchill, C.~W., Kacprzak, G.~G., \& Murphy, M.~T.\ 2013a, \apj, 776, 114 
\bibitem[Nielsen et al.(2013b)]{NielsenMgIIcatalogII} Nielsen, N.~M., Churchill, C.~W., \& Kacprzak, G.~G.\ 2013b, \apj, 776, 115 
\bibitem[P{\^a}ris et al.(2012)]{Parisdr9qso} P{\^a}ris, I., Petitjean, P., Aubourg, {\'E}., et al.\ 2012, \aap, 548, A66 
\bibitem[Peek et al.(2014)]{Peek2014} Peek, J., M\'enard, B., Corrales, L., submitted.
\bibitem[Rao et al.(2006)]{2006ApJ...636..610R} Rao, S.~M., Turnshek, D.~A., \& Nestor, D.~B.\ 2006, \apj, 636, 610 
\bibitem[Rubin et al.(2013)]{Rubin2013} Rubin, K.~H.~R., Prochaska, J.~X., Koo, D.~C., et al.\ 2013, arXiv:1307.1476
\bibitem[Schneider et al.(2010)]{Schneiderdr7qso} Schneider, D.~P., Richards, G.~T., Hall, P.~B., et al.\ 2010, \aj, 139, 2360
\bibitem[Smith et al.(2007)]{SmithPAH} Smith, J.~D.~T., Draine, B.~T., Dale, D.~A., et al.\ 2007, \apj, 656, 770 
\bibitem[Steidel et al.(1994)]{Steidel1994} Steidel, C.~C., Dickinson, M., \& Persson, S.~E.\ 1994, \apjl, 437, L75 
\bibitem[Steidel et al.(2010)]{2010ApJ...717..289S} Steidel, C.~C., Erb, D.~K., Shapley, A.~E., et al.\ 2010, \apj, 717, 289 
\bibitem[Tumlinson et al.(2011)]{TumlinsonOVI} Tumlinson, J., Thom, C., Werk, J.~K., et al.\ 2011, Science, 334, 948
%\bibitem[Tumlinson et al.(2013)]{TumlinsonHI} Tumlinson, J., Thom, C., Werk, J.~K., et al.\ 2013, \apj, 777, 59 
\bibitem[Weiner et al.(2009)]{Weiner2009} Weiner, B.~J., Coil, A.~L., Prochaska, J.~X., et al.\ 2009, \apj, 692, 187
%\bibitem[Werk et al.(2013)]{Werk2013} Werk, J.~K., Prochaska, J.~X., Thom, C., et al.\ 2013, \apjs, 204, 17 
\bibitem[Werk et al.(2014)]{Werk2014} Werk, J.~K., Prochaska, J.~X., Tumlinson, J., et al.\ 2014, \apj, 792, 8
\bibitem[Wright et al.(2010)]{WrightWISE} Wright, E.~L., Eisenhardt, P.~R.~M., Mainzer, A.~K., et al.\ 2010, \aj, 140, 1868
\bibitem[York et al.(2000)]{YorkSDSS} York, D.~G., Adelman, J., Anderson, J.~E., Jr., et al.\ 2000, \aj, 120, 1579
%Z
\bibitem[Zhu \& M{\'e}nard(2013)]{ZhuMgIIcatalog} Zhu, G., \& M{\'e}nard, B.\ 2013, \apj, 770, 130
\bibitem[Zhu \& M{\'e}nard(2013)]{ZhuCaII} Zhu, G., \& M{\'e}nard, B.\ 2013, \apj, 773, 16 
\bibitem[Zhu et al.(2014)]{ZhuMgII} Zhu, G., M{\'e}nard, B., Bizyaev, D., et al.\ 2014, \mnras, 439, 3139
\bibitem[Zibetti et al.(2007)]{Zibetti2007} Zibetti, S., M{\'e}nard, B., Nestor, D.~B., et al.\ 2007, \apj, 658, 161


\end{thebibliography}
\end{document}